\documentclass[preprint,journal]{vgtc}       

\ifpdf
  \pdfoutput=1\relax                   
  \usepackage{graphicx}                
  \DeclareGraphicsExtensions{.pdf,.png,.jpg,.jpeg} 
\else
  \usepackage{graphicx}                
  \DeclareGraphicsExtensions{.eps}     
\fi%

\graphicspath{{figures/}{pictures/}{images/}{./}} 

\usepackage{microtype}                 
\PassOptionsToPackage{warn}{textcomp}  
\usepackage{textcomp}                  
\usepackage{mathptmx}                  
\usepackage{times}                     
\usepackage{cite}                      
\usepackage{tabu}                      
\usepackage{booktabs}                  

\newcounter{code}
\makeatletter
\newcommand*{\codeLabel}{%
  \@dblarg\@codeLabel
}
\def\@codeLabel[#1]#2{%
  \begingroup
    \renewcommand*{\thecode}{\textbf{(C\arabic{code})}}%
    \refstepcounter{code}%
    \label{#1}%
    \vspace{1ex}

    \noindent\textbf{C\arabic{code}. #2}%
  \endgroup
}
\makeatother

\makeatletter
\newcommand\footnoteref[1]{\protected@xdef\@thefnmark{\ref{#1}}\@footnotemark}
\makeatother

\usepackage[dvipsnames]{xcolor}

\definecolor{cbSurveyRed}{RGB}{228,26,28}
\definecolor{cbSurveyBlue}{RGB}{55,126,184}
\definecolor{cbSurveyGreen}{RGB}{77,175,74}
\definecolor{cbSurveyPurple}{RGB}{152,78,163}
\definecolor{cbSurveyOrange}{RGB}{255,127,0}

\definecolor{cbTeaserTeal}{RGB}{27,158,119}
\definecolor{cbTeaserOrange}{RGB}{217,95,2}
\definecolor{cbTeaserPurple}{RGB}{117,112,179}
\definecolor{cbTeaserPink}{RGB}{231,41,138}

\onlineid{1251}

\vgtccategory{Research}
\vgtcpapertype{Empirical Study}

\title{Guidelines For Pursuing and Revealing Data Abstractions}

\author{Alex Bigelow, Katy Williams, and Katherine E. Isaacs}
\authorfooter{
\item
 Alex Bigelow, Katy Williams, and Katherine E. Isaacs are with the University of Arizona. E-mail: {alexrbigelow@email | kawilliams@email | kisaacs@cs}.arizona.edu.
}

\shortauthortitle{Bigelow \MakeLowercase{\textit{et al.}}: Guidelines For Pursuing Latent Data Abstractions}

\abstract{
Many data abstraction types, such as networks or set relationships, remain unfamiliar to data workers beyond the visualization research community. We conduct a survey and series of interviews about how people describe their data, either directly or indirectly. We refer to the latter as latent data abstractions. We conduct a Grounded Theory analysis that (1) interprets the extent to which latent data abstractions exist, (2) reveals the far-reaching effects that the interventionist pursuit of such abstractions can have on data workers, (3) describes why and when data workers may resist such explorations, and (4) suggests how to take advantage of opportunities and mitigate risks through transparency about visualization research perspectives and agendas. We then use the themes and codes discovered in the Grounded Theory analysis to develop guidelines for data abstraction in visualization projects. To continue the discussion, we make our dataset open along with a visual interface for further exploration.
} 


\teaser{
  \center{\includegraphics[width=\textwidth]{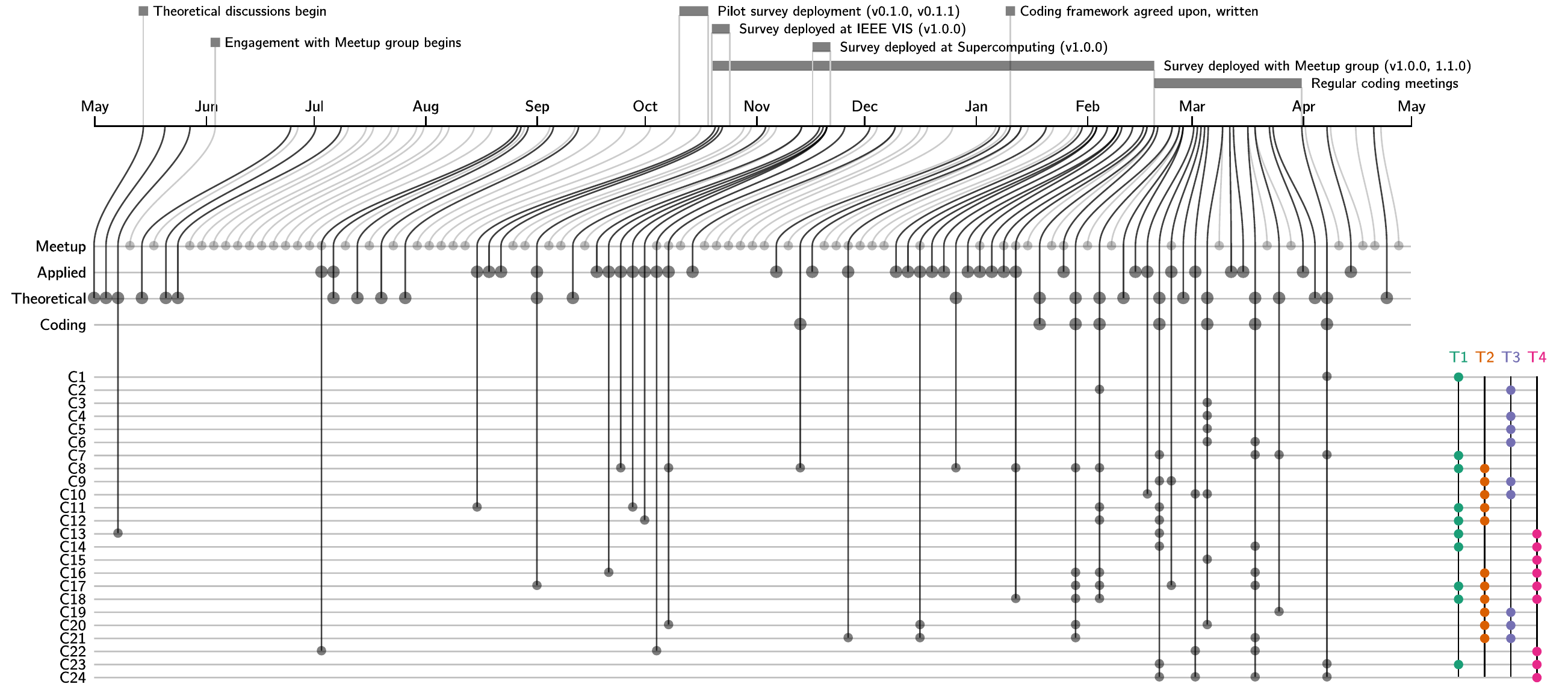}}
  \caption{\label{fig:lattice} A summary of study events over time, their
  temporal relationship with memos, memo relationships with codes, and code
  relationships with themes. The timeline at the top shows the timing of study
  events, with curved lines indicating when individual memos were created. The
  four rows below the timeline indicate the nature of the context in which memos
  were written, including Meetup attendance, when data workers discussed their
  applied datasets, when the authors engaged in theoretical discussions, and
  when the authors engaged in open coding. Rows C1-C24 show which memos directly
  informed the development of codes. Columns
  \textbf{\textcolor{cbTeaserTeal}{T1}}--\textbf{\textcolor{cbTeaserPink}{T4}}
  show which codes directly inform which themes.}
}

\vgtcinsertpkg

\begin{document}


\firstsection{Introduction}

\maketitle

Data abstractions are fundamental to a wide set of visualization activities,
from performing and documenting the provenance of data wrangling operations,
to understanding the mental models of domain experts in design study research,
to justifying design decisions in technique- or systems-focused research, and
to reasoning about the role of data abstraction in theoretical visualization
research. Difficulties in reasoning about and communicating data abstractions
therefore have far-reaching implications: effective communication about data
abstractions is critically important to the way researchers justify design
decisions in technique- or systems-focused research. A poor understanding of
the mental models of domain experts in design study research is a significant
threat that risks creating solutions and systems that do not address real
needs~\cite{munzner_NestedModel_2009}. Too much focus on a single data
abstraction has been observed to limit
creativity~\cite{bigelow_ReflectionsHow_2014} and to warp scientific
analysis~\cite{bigelow_DrivingGenetics_2012}. However, the extent to which
these effects apply, in terms of specific abstractions, is poorly
understood.

We set out to understand how malleable a data abstraction is, and to better understand the process of pursuing \textbf{latent data abstractions}. We define a latent data abstraction to be one that is meaningful
and useful, yet undiscovered. It has yet to be fully elucidated, communicated,
documented, and formatted. A data abstraction becomes less latent as coherent
details are identified, as its details are spoken or written, and as its
artifacts in a computer are actualized into relevant forms.

Because there were blind spots in the questions that we should even ask, we
chose to conduct a Grounded Theory Method investigation seeking to discover how
a diverse range of \textbf{data workers}, from spreadsheet users to programmers,
across different disciplines, consider different data abstractions. This
investigation analyzes \textbf{memos}, or research field notes taken during conversations, meetings, and interviews, as well as the results of a deployed survey.

The result is an evidence-based set of codes and themes regarding
data abstractions with implications for how project teams and
individuals discover, wrangle, manage, and report their data abstractions. In
particular, we find that introducing a \textbf{data abstraction typology}---a
model that describes the space of possible data abstractions and/or data
wrangling operations---can elicit rich communication and reflection about data
and uncover latent data abstractions, even when such a typology is imperfect. We
show how visualization researchers can increase actionable communication with
data workers by introducing and critiquing a typology together, as a
visualization design activity~\cite{mckenna_DesignActivity_2014a}.

The codes and themes in this paper also add to existing literature by
explaining some of the reasons why communicating about data abstractions can be
so challenging. Reflecting on these themes and our collective interactions with data workers, we provide guidelines for communicating with data workers
about data abstractions, that also have applications for more
crisp communication about data abstractions in design study, technique, systems,
and theoretical research papers.

We have made the raw data collected in our survey available through an
interactive visual interface.\footnote{\label{foot:surveyData}osf.io archive of
survey responses: \url{https://osf.io/s2wmp/} \\ \hspace*{1.75em}Selected
response visualizations are included in the supplemental material} We also
include a version-controlled archive\footnote{\label{foot:codesRepo}osf.io
archive of codes, themes, and audit: \url{https://osf.io/382fn/}} of codes,
themes, and an audit trail~\cite{carcary_ResearchAudit_2009} that summarize
memos of observations from a year of interviews and meetings with diverse data
workers, as well observations from the visualized survey responses.

In summary, our contributions are:

\vspace{-1.5ex}

\begin{enumerate}
  \itemsep=0.2ex
  \item A set of themes, supported by codes, that describe phenomena associated with data abstractions that arise in the processes of visualization design and data wrangling (\autoref{sec:theory}),

  \item Guidelines for developing data abstractions
    (\autoref{sec:guidelines}),

  \item The design of an open survey regarding the description of data and the
    malleability of data abstractions (\autoref{sec:survey},
    \autoref{sec:reflections}), and

  \item An open, visualized corpus of survey responses.\footnoteref{foot:surveyData}

\end{enumerate}

\vspace{-1.5ex}

We begin by discussing necessary background and a review of related work
(\autoref{sec:related}), and our methodology
(\autoref{sec:methodology}). We present the codes derived from
our study (\autoref{sec:codes}) and how they come together to form themes (\autoref{sec:theory}). We follow with guidelines and reflections (\autoref{sec:discussion}).

\section{Background and Related Work}
\label{sec:related}

We discuss the theoretical underpinnings of our work
(\autoref{sec:underpinnings}), related background in thinking and communicating about data in
analysis and design projects (\autoref{sec:relatedcomm}), the importance of
documenting real-world wrangling needs (\autoref{sec:dataAboutWrangling}), and the context in which this work fits into research into creativity (\autoref{sec:creativeroles}).

\subsection{Theoretical Underpinnings}
\label{sec:underpinnings}

This study employs a team-based~\cite{wiener_MakingTeams_2007}, interpretivist
form~\cite{weed_CapturingEssence_2017} of Grounded Theory Methodology, resulting
in the development and refinement of four themes---these four themes, with their
supporting codes, comprise what is often termed a {\em substantive
theory}~\cite{muller_CuriosityCreativity_2014}.

Grounded Theory is an approach that is uniquely suited for investigating and
describing phenomena in which questions evolve rapidly. The general pattern of a
Grounded Theory investigation involves identifying and refining \textbf{codes},
or concepts that describe phenomena while conducting diverse research
activities, such as performing interviews or conducting a survey. The choice of
research activity is typically informed by the codes as questions evolve. As
codes mature and are grouped into categories, they begin to form
\textbf{themes}, or evidence-based hypotheses about a phenomenon. Eventually,
codes and themes reach \textbf{saturation}, or a point at which researchers are
confident that codes and themes are stable and no additional data needs to be
collected.

Grounded Theory Methodology was an appropriate fit for beginning this
investigation because our initial suspicions---that non-tabular data
abstractions may be comprehensible, useful, and under-utilized among the broad
population of data workers---were very general and based on a small number of
surprising
observations~\cite{bigelow_DrivingGenetics_2012,bigelow_ReflectionsHow_2014}.
The nature of the questions that we should pursue were prone to rapid revision
and refinement as additional, surprising observations arose.

There are many ways to conduct a Grounded Theory investigation. In this study,
we identified, discussed, and refined each code and theme as a
team~\cite{wiener_MakingTeams_2007}. We used \emph{surprise} as a principled way
to guide our choice of research
activities~\cite{muller_CuriosityCreativity_2014}; the extent to which we
pursued interactions with data workers, and adapted and deployed a survey, were
motivated by identifying gaps in our own knowledge and unanticipated findings.
In contrast, we also used our lack of surprise as a qualitative indicator to
know when codes and themes had reached saturation.

Grounded Theory can also be employed for different epistemological goals. In
contrast to the positivist research that we typically see in the visualization
research community~\cite{meyer_CriteriaRigor_2019}, our Grounded Theory
investigation had interpretivist objectives~\cite{weed_CapturingEssence_2017}.
\textbf{Interpretivist} research aims to {\em describe} phenomena and {\em
generate} hypotheses. This is in contrast to the \textbf{positivist} approach
used in the scientific method that aims to {\em test} hypotheses. The four
interpretivist themes that we identify, and their supporting codes, are
transferable, in contrast to the way that formal theories are generalizable.
Both intellectual traditions require systematic analysis of evidence, but the
nature of supporting data and the ways that data are collected and analyzed are
different.

In presenting qualitative research, we are careful of
pitfalls~\cite{sandelowski_RealQualitative_2001} in reporting numbers and
counts: we include the visualized corpus of survey
responses\footnoteref{foot:surveyData} to maximize available context. Our
numeric statements and visualizations are meant as interpretivist
descriptions of phenomena associated with how data workers think and
communicate about data abstractions, not positivist statements of statistical
significance.

Although this is not a visualization design study, Meyer and Dykes' six
categories for judging and reporting rigor~\cite{meyer_CriteriaRigor_2019} are
relevant for the kind of interpretivist research that we present. This research
is \emph{informed} by our relevant prior research experiences; \emph{reflexive}
in our efforts to \emph{constantly
compare}~\cite{charmaz_ConstructingGrounded_2014} collected data and gaps in our
understanding; \emph{abundant} through the number of survey participants and
diversity of interview and Meetup participants; \emph{plausible} through
documented connections from memos and survey responses, to codes, and to themes;
\emph{resonant} in that the themes have broad implications for how
visualization research is conducted and reported; and \emph{transparent} through
the public release of the survey, its responses, and the revision history of the
evolution of our codes, themes, and relevant metadata.

\subsection{Thinking and Communicating About Data}
\label{sec:relatedcomm}

We build on other efforts to understand how data workers think and communicate about data. From the beginning of our research, our main focus has been to expand understanding of one specific approach identified by Muller \textit{et al.}~\cite{muller_HowData_2019}: how data workers approach the \emph{design} of their data, as opposed to \emph{discovery}, \emph{capture}, \emph{curation}, and \emph{creation}.

Many authors have noted the designed nature of data abstractions~\cite{meyer_NestedBlocks_2015}, such as the handcrafted nature of many cybersecurity datasets~\cite{kiss_EvaluationManually_2013}. Feinberg observes that the mere use of a dataset makes the user a designer of its abstraction~\cite{feinberg_DesignPerspective_2017}, even if users are unaware of their inherent flexibility. Consequently, there is a need to learn to develop a ``data vision'' to exercise discretion and creativity in designing abstractions~\cite{passi_DataVision_2017}. This is especially important in light of ethical responsibilities to structure data effectively~\cite{dagandra_InformFormInformation_2012}, as the design of what is measured and how it is stored can be overtly political acts~\cite{pine_PoliticsMeasurement_2015}.

The responsibility to design effective abstractions does not always fall upon data workers in isolation. In the context of
visualization design studies that involve individuals with diverse roles and
expertise, effective data abstraction design~\cite{munzner_NestedModel_2009}
and communication about abstractions as they
evolve~\cite{sedlmair_DesignStudy_2012}, are critical to the success of a
project.

However, difficulties arise in effectively communicating about data
abstractions~\cite{pretorius_WhatDoes_2009}. There are myriad aspects to data
abstractions in design projects, such as adapting to data changes, anticipating
edge cases, understanding technical constraints, articulating data-dependent
interactions, communicating data mappings, and preserving data mapping integrity
across iterations~\cite{walny_DataChanges_2019}. These difficulties are
consistent with reports of there being surprisingly little documentation about
the design of abstractions~\cite{zhang_HowData_2020}. The lack of documentation
makes human decisions invisible and threatens future analysis. In strictly
machine-learning contexts, some authors have gone as far as suggesting that
``deemphasizing the need to understand algorithms and
models''~\cite{passi_TrustData_2018} may be an effective way to increase trust
in model predictions. We show that the inverse is also true: that education and
transparency can foster healthy skepticism of data models and abstractions,
which can be important for fairness and provenance. We argue that
transparency about data abstractions can be especially important for data
wrangling and visualization, in which data workers need to ``interact not only
with the interface but \emph{with the data}''~\cite{walny_DataChanges_2019}.

To facilitate communication about a particular project's specific data
abstraction, the visualization research community often relies extensively upon
data abstraction
typologies~\cite{munzner_WhatData_2014,figueiras_TypologyData_2013,chi_TaxonomyVisualization_2000}.
Currently, the main purposes of such typologies are to guide a researcher in the
selection of appropriate visual encodings, and to support transferability across
different design studies. However, aside from highly contextual design study
research itself, there is little data that reveals the extent to which the
visualization research community's typologies are compatible with data workers'
perspectives and language, and, although the interventionist nature of design
study research is known~\cite{mccurdy_ActionDesign_2016a}, the effects of
introducing foreign data concepts have yet to be described in detail.

\subsection{Data about Applied Wrangling Needs}
\label{sec:dataAboutWrangling}

Little applied data wrangling work has been published in the visualization
community, even though novel algorithms, data structures, and infrastructure
need to be implemented in ways that correctly address nuanced worker needs. Such
efforts often consume the bulk of the labor involved in applied visualization
research~\cite{guo_ProactiveWrangling_2011,kandel_WranglerInteractive_2011,muller_HowData_2019},
and can include rich refinements in terms of task clarity and data location that
advance science and constitute important visualization research contributions in
their own right~\cite{sedlmair_DesignStudy_2012}, yet, without also engineering
a polished visualization system, such work has lacked clear publication venues.

The lack of such work leaves a major gap in needed visualization research.
Although our work includes a qualitative dataset that only begins to fill this
gap, the extent to which data workers use or even consider different data
abstractions is still difficult to analyze or test, as data wrangling decisions
are rarely documented in research or in practice~\cite{zhang_HowData_2020}. When
such decisions are documented in research, they typically only exist as
justification for a visualization design; resulting in limited information about
the data abstraction, its provenance, and important documentation about how and
why it was reshaped.

It can consequently be difficult to justify technique-driven or systems-focused
research into general-purpose data wrangling software
systems~\cite{guo_ProactiveWrangling_2011,kandel_WranglerInteractive_2011,verborgh_UsingOpenRefine_2013,heer_OrionSystem_2011a,liu_PloceusModeling_2014,srinivasan_GraphitiInteractive_2018a,bigelow_OrigraphInteractive_2019},
as such efforts often lack grounding in real user needs. Instead, they are
forced to rely upon past researcher experience, scant hints about real-world
data wrangling precedents that exist in design study literature, and speculation
about how data workers might think and what operations they might find useful. This study, and future standalone publications that are focused on data transformations, can help to better inform the design of such systems.


\subsection{Creativity and Creative Roles}
\label{sec:creativeroles}

Discovering a latent data abstraction can have powerful creative benefits, such
as inspiring radical visual
innovations~\cite{nielsen_ABySSExplorerVisualizing_2009,mckenna_DesignActivity_2017a}.
Although the work that we present has implications for visualization researchers
and their interactions with the broader population of data workers, our primary
objective is to compare and contrast sets of creative objectives that can be
held by any kind of data worker---including visualization researchers
themselves. Consequently, we identify the role of an \textbf{abstraction
theorist} that seeks to discover useful latent data abstractions, and contrast
that objective against the broad set of all other concerns that a data worker
may need to consider, such as data wrangling, data ownership, workflow
management, the design and implementation of visualizations, evaluation, and reporting on visualization research.

Contrasting these roles is similar in spirit to Von Oech's popularized
``explorer, artist, judge, warrior'' creative
roles~\cite{vonoech_KickSeat_1986}: ``theorist'' and ``worker'' may refer to
distinct individuals in a collaborative environment, such as a visualization
researcher and domain expert, or they could refer to different priorities that a
single individual is considering on their own. Therefore, we describe
differences through a pragmatic lens, instead of analyzing different
populations' creative styles or
cognition~\cite{sternberg_HandbookCreativity_1999}.

We add to precedents for pragmatic guidance for creativity in visualization
design, including creativity
workshops~\cite{goodwin_CreativeUserCentered_2013,kerzner_FrameworkCreative_2019}
and exercises~\cite{mckenna_DesignActivity_2014a}---we propose the
pursuit of latent data abstractions as an additional creativity exercise,
specific to the design of data itself.

\section{Methodology}
\label{sec:methodology}

The evidence upon which we base our findings comes from two sources: memos and a deployed survey about data abstraction perspectives. It is important to note that, consistent with our interpretivist objectives, many of the following methods are deliberately \textit{uncontrolled}---rather than testing hypotheses, our goal is to ask better questions. Here we discuss both sources of data, and the way that they both influenced, and were influenced by, our internal data abstraction typology.

\subsection{Memos and Timeline}

We wrote memos in four contexts: 1) regular attendance at data-centered
community Meetups, 2) applied conversations with data workers in diverse
contexts about their perspective on their data, 3) theoretical discussions
about data abstractions among the authors, and 4) collaborative open coding
sessions. A summary of all memos, their relationships with codes, and code
relationships with themes, are shown in \autoref{fig:lattice}, and an
associated audit trail~\cite{carcary_ResearchAudit_2009} is in the
supplemental material.\footnoteref{foot:codesRepo}

This project began with theoretical conversations about the nature of data
abstractions between the authors, that arose occasionally as part of regular
meetings. Early on, we decided to engage with an existing local Meetup group
that regularly met to seek or provide help with data: a core group of regular
members met twice per week at a coffee shop or bar, and continued to meet
remotely beginning in March due to social distancing measures. Members and
visitors frequently brought laptops to show data and code that they were working
with, to solicit advice or help with debugging in a casual context. The core
group and its frequent visitors included a diverse array of researchers,
administrators, and data scientists from the local university and surrounding
community. As these meetings and interactions were largely \textit{ad-hoc}, an
accurate count of all informants is impossible to report, however, a
selected subset of these community members---those that provided specific
information that informed the development of a code---are shown in
Table~\ref{table:informants}.

Later, as our survey was developed, it was deployed among this group, as well as
at the 2019 IEEE VIS and 2019 Supercomputing conferences. Each of the 219 survey
responses are included in the supplemental
material.\footnoteref{foot:surveyData} Deployments of the survey often prompted
conversations that provided additional valuable insight that we added to our
growing set of memos.

As concepts and patterns began to be less surprising, the authors began to
identify codes from supporting evidence, in a collaborative open coding
environment similar to the one described by
Wiener~\cite{wiener_MakingTeams_2007}. After writing and agreeing upon a
framework for documenting codes in a version-controlled
repository\footnoteref{foot:codesRepo}, the authors began to meet 2-3 times per
week to discuss, refine, and write codes that we had identified as we reviewed
survey responses and our individual field notes. As we discussed different
patterns in the data, each author actively
cited~\cite{moravcsik_ActiveCitation_2010} supporting personal experience, memos
from a related interview, or specific survey responses to support or contest the
proposed code. Where personal experience was identified as evidence, additional
memos were written to document the experience. As we began to observe broader
themes across codes, these were also written, discussed, refined, and connected
to codes. Refinements to codes included citing additional evidence, rephrasing
codes, splitting codes, or combining codes in an \textit{ad-hoc} process similar
to affinity
diagramming~\cite{beyer_ContextualDesign_1999,hanington_UniversalMethods_2012},
but in a version-controlled text file instead of using cards or notes. Finally,
an audit was conducted to verify the nature of the source data, the
relationships between memos and survey responses to codes, and the relationships
between codes and themes; the result of the audit is visualized in
\autoref{fig:lattice}.

\begin{table}[!b]
  \vspace{-0.8em}
  \caption{Informants}
  \label{table:informants}
\begin{tabular}{@{}lll@{}}
\toprule
Informant & Role                    & Domain                      \\ \midrule
I1      & Professor               & Medicine / Bioengineering   \\
I2      & Research Assistant      & Linguistics                 \\
I3      & Postdoctoral Researcher & Biology                     \\
I4      & Research Director       & Information technology      \\
I5      & Professor               & Mathematics                 \\
I6      & Research Director       & Interdisciplinary institute \\
I7      & Professor               & Interdisciplinary institute \\
I8      & Postdoctoral Researcher & Information science         \\
I9      & Postdoctoral Researcher & Biology                     \\
I10      & Postdoctoral Researcher & Biology                     \\
I11      & Program Coordinator     & Interdisciplinary institute \\
I12      & Postdoctoral Researcher & Bioinformatics              \\
I13      & Professor               & Public health               \\
I14      & Professor               & Biology                     \\
I15      & Professor               & Computer Science            \\
I16      & Professor               & Computer Science            \\
I17      & Research Assistant      & Computer Science            \\
I18      & Engineer                & Industry                    \\
I19      & Data Scientist          & Industry                    \\ \bottomrule
\end{tabular}
\end{table}

\subsection{Data Abstraction Typology Evolution}
\begin{figure}[!t]
  \center{\includegraphics[width=\linewidth]{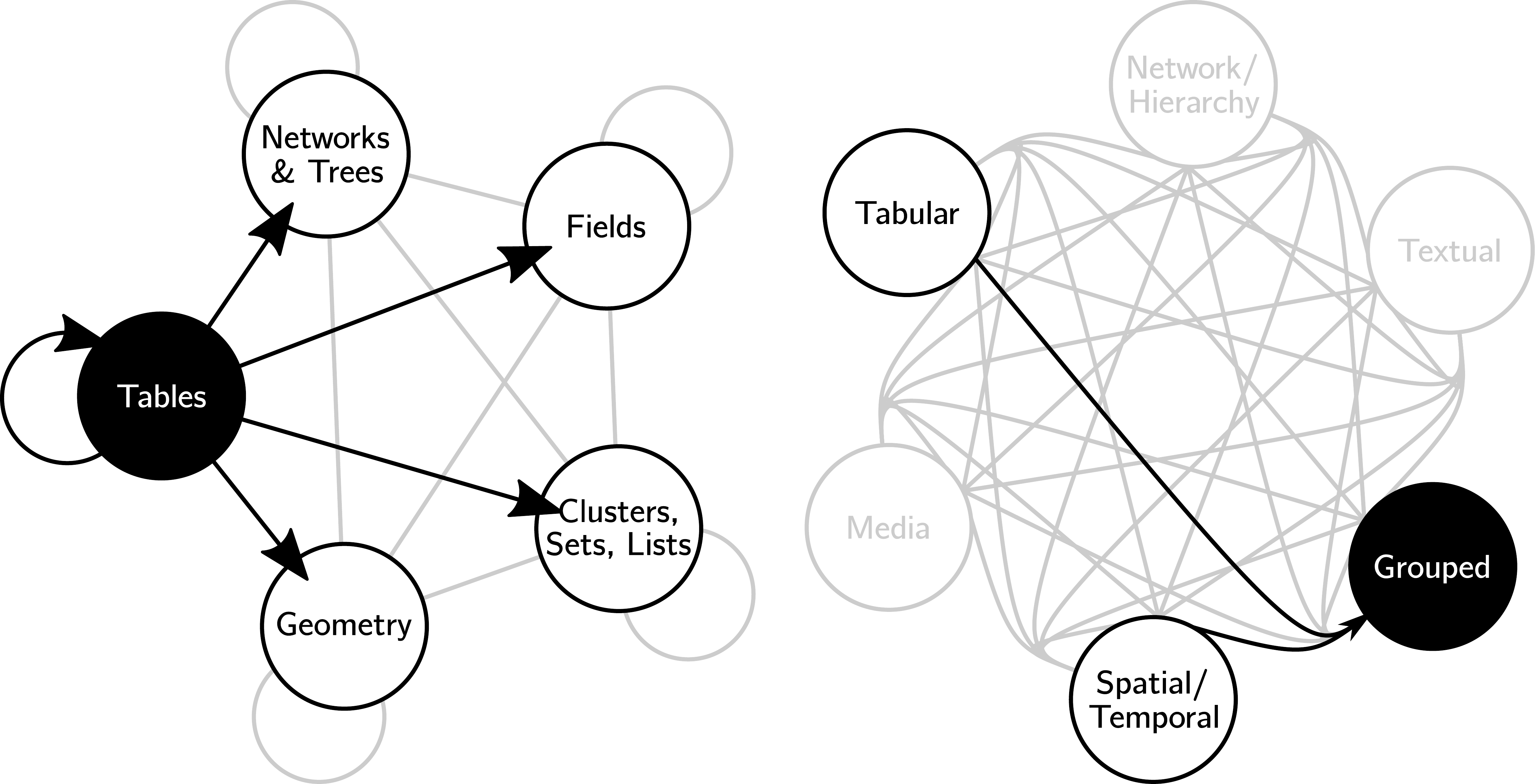}}
  \caption{\label{fig:typologyEvolution} The evolution of our data abstraction
  typology. Initially, we modeled abstractions as fitting into five specific
  data abstraction types, with every node in the complete graph representing a
  potential latent abstraction (left). Data wrangling operations, such as
  converting rows in a table as nodes in a network, or performing dimensionality
  reduction of tabular columns in a high-dimensional space, are modeled as
  directed edges that require changing to a non-tabular data abstraction. As we
  engaged in applied conversations with data workers and designed our survey,
  the specific categories in our typology evolved, as did its model. The final
  typology models the process of considering a latent abstraction as a hyperedge
  coming from a hybrid set of different categories to a new target latent
  abstraction (right), such as imagining ways to cluster rows in a table based
  on columns containing geographic information.}
  \vspace{-1em}
\end{figure}

We began our investigation by adapting the data abstraction typology described
by Tamara Munzner~\cite{munzner_WhatData_2014} to a data wrangling context: our
initial objective was to describe a design space of possible data wrangling
operations, so we modeled operations as edges in a complete graph, connecting
each of five broad data abstraction types, as shown on the left in
\autoref{fig:typologyEvolution}. Each edge represents a transition from one
data type to another; for example, network modeling
tools~\cite{heer_OrionSystem_2011a,liu_PloceusModeling_2014,srinivasan_GraphitiInteractive_2018a,bigelow_OrigraphInteractive_2019}
would largely support operations along an edge from ``Tables'' to ``Networks \&
Trees.'' Self-edges describe wrangling operations that transform a dataset to a
different form of the same type, such as transposing the rows and columns of a
table.

We quickly discovered many weaknesses of this model, through our own theoretical
discussions and applied conversations with data workers. Most datasets have
elements or relevant metadata that could be described heterogeneously: with more
than one type of abstraction. Furthermore, many of the dataset types that we
initially selected were a poor fit for specific datasets, such as text corpora.
These weaknesses caused us to reflect on the overall purpose of such a model:
one that attempts to delineate all possible data wrangling operations may not be
possible. However, adapting it to be more flexible could potentially aid in the
process of exploring latent data abstractions.

Motivated by the weaknesses that we had discovered about our initial model, we
changed the focus of our investigation from modeling the space of possible data
wrangling operations to investigating how malleable a set of identified
abstraction types can be in practice, and the extent to which enforcing a
different perspective on a real-world dataset can have creative benefits in its
own right. We pivoted from attempting to develop a model, to developing and
deploying a survey. Consequently, we do not present our preliminary data
abstraction typology as a contribution, even though it guided the development of
the survey and it informs our codes and themes.

For our survey, we adapted our model to describe the act of theorizing about an
alternative data abstraction, instead of performing a concrete data wrangling
operation---although it could still reveal unmet needs for data wrangling tools,
that objective was no longer prioritized. We modeled these acts as hyperedges,
also shown in the right side of \autoref{fig:typologyEvolution}, with the target
alternative abstraction remaining singular but allowing for any combination of
source abstractions. This makes it possible for survey participants to describe
their dataset with any combination of abstraction types, and yet still explore
an abstraction type that may be less familiar.

\begin{figure*}[!h]
  \center{\includegraphics[width=\textwidth]{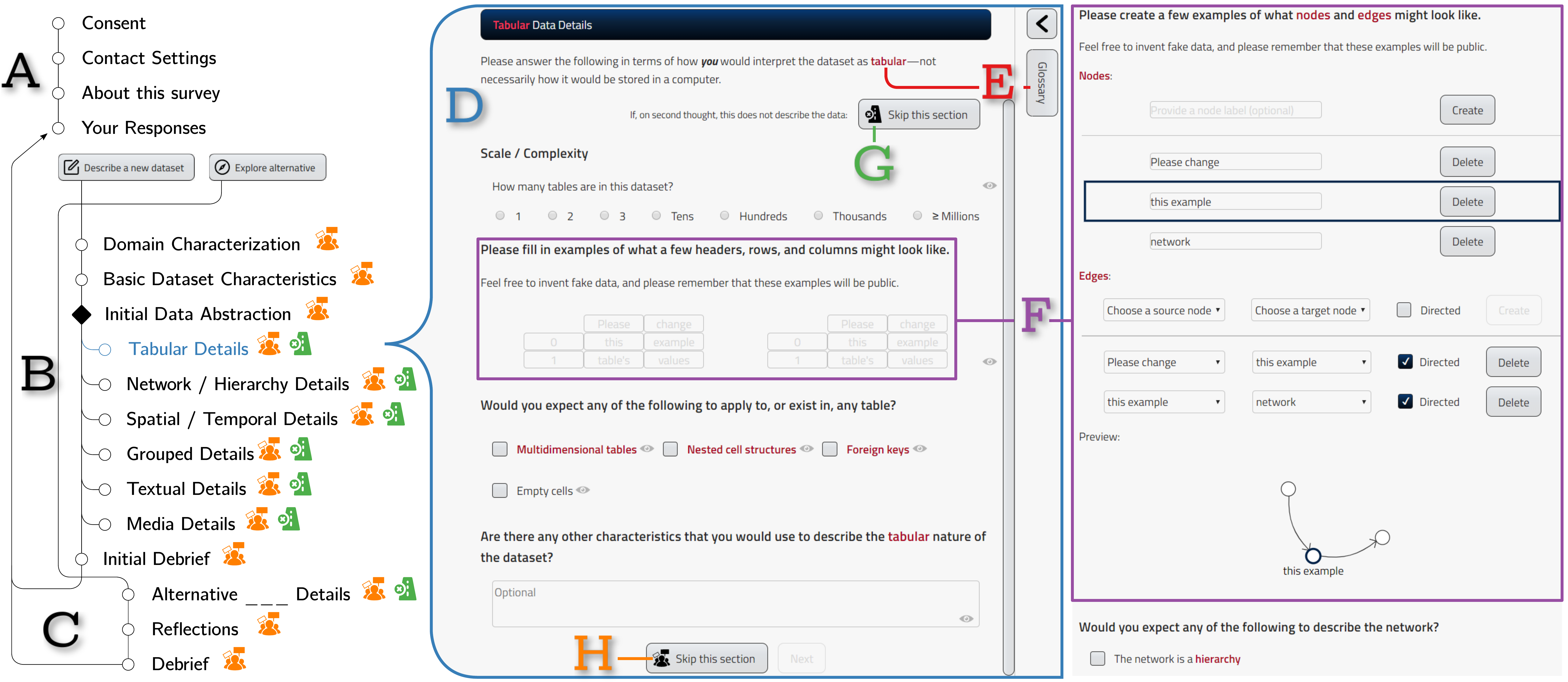}}
  \caption{\label{fig:survey} An overview of the survey that we deployed. The survey is divided into three sections, shown here as a flow diagram. The first section (A) includes consent forms, contact settings, an introduction to the innovations in the survey, and a summary of responses that redirect to the other two survey portions. The main ``Describe a new dataset'' portion of the survey (B) invites participants to describe a real or imagined dataset, and asks them to reflect upon the extent to which they think about the dataset in terms of the six dataset types that we identified. Where participants reply that they at least ``rarely'' think of their data in terms of a given type, they are asked for more details in a specialized Details section of the survey. The final ``Explore alternative'' portion of the survey (C) invites participants to imagine their dataset as the type that they initially thought about the \emph{least}, and fill in the associated Details portion of the survey with this new perspective. As an example, the Tabular Details interface is shown \textcolor{cbSurveyBlue}{(D)}. Participants are encouraged throughout the survey to look up terminology highlighted in red, where participants can edit the terms and suggest alternative definitions in the glossary \textcolor{cbSurveyRed}{(E)}. In some Details sections, participants are asked for a small sample of what they imagine the data to look like, to help ground their thinking \textcolor{cbSurveyPurple}{(F)}. At any point in a Details section \textcolor{cbSurveyGreen}{(G)}, or at the end of most other sections \textcolor{cbSurveyOrange}{(H)}, participants can choose to skip the section to provide targeted critique on the survey itself if the questions have strayed far enough from the participant's mental model.}
  \vspace{-0.8em}
\end{figure*}

\subsection{Open-ended Survey Design and Deployment}
\label{sec:survey}

We developed and deployed our survey in the form of an interactive web page. It
is designed in three phases, shown in \autoref{fig:survey}: after the first
introductory phase, the main phase of the survey invites the participant to
describe a real or imagined dataset, in terms of our data abstraction typology's
six broad data abstraction categories: tabular data, network / hierarchical
data, spatial / temporal data, grouped data, textual data, or media. Further
details are requested from participants where they indicate that they at least
``rarely'' interpret the data in terms of a particular abstraction category.

The final phase of the survey chooses randomly from the abstractions that a
participant has indicated that they think about the \emph{least}, and encourages
them to try to think creatively about their data with that abstraction. It
solicits qualitative, self-reported feedback on the extent to which the imagined
transition would be useful, as well as which software tools
participants would likely use to accomplish the transition.

Throughout all phases, the survey solicits meta-feedback about the survey
itself. Participants can challenge the survey design, questions asked, our set
of possible data abstraction categories, and the terminology that we used. The
glossary is interactive, allowing participants to provide alternate terms or
definitions. Participants can also skip sections of the survey that begin to ask
questions that the participant feels have strayed from their perspective or use
case.

We deployed the survey among three groups: attendees at regular community data
Meetups, attendees at the 2019 IEEE VIS Conference, and attendees at the 2019
Supercomputing Conference. Although the latter two groups had a less diverse
computing focus, we were aware of ongoing discussions about data abstractions
within these communities, and suspected that these groups were particularly
likely to offer direct critiques of our typology and approach.

\section{Codes}
\label{sec:codes}

Here we present the codes for phenomena that we identified in our collaborative open coding
process, with selected supporting evidence. For more detailed supporting
evidence for each code, see the supplemental
archive.\footnoteref{foot:codesRepo} As there are 24 codes, we present them as groups for readability, however, the analyzed themes presented in \autoref{sec:theory}, and their relationships with codes, are more complex, as shown in \autoref{fig:lattice}.

\vspace{1em}
\hrule
\vspace{1ex}

\noindent Codes \textbf{C1--C6} are based mostly on patterns that we observed in
the visualized corpus of survey responses.

\codeLabel[code:DISCREPANCY_BETWEEN_MENTAL_MODELS_AND_RAW_DATA]{Compared to the
diverse responses in how participants described thinking about their data, the
way that they characterized how it is represented in a computer was
disproportionately tabular.} This disconnect between the mental model and
physical computer representation indicates not only a possible need for new data
storage or data wrangling tools but also a lack of awareness of other data
storage options. Data workers may default to tabular data organization because
it more easily fits into their current workflow and tools, or because they do
not know of existing ``unconventional,'' non-tabular tools.

\codeLabel[code:RESPONSE_SIZES]{There was wide variation in reported dataset
scales.} Taken from the median response for each of the ``Basic Dataset
Characteristics'' questions (e.g. ``Approximately how large is this
dataset?''), the median dataset was on the order of megabytes (close to
gigabytes) in size, with thousands of items in the dataset and tens of
attributes.

\codeLabel[code:REFERENCED_TECHNIQUES_INSTEAD_OF_TOOLS]{Participants included
broad techniques in their responses for wrangling tool support.} When asked to
actually transform their initial dataset into the alternative abstraction type,
most participants listed software tools or programming languages but some listed
techniques. These techniques included natural language processing (``NLP,
Python'', ``Python, nlp techniques''), machine learning, and mathematical
operations (``cluster into connected components'', ``Morse Smale Complex'').

\codeLabel[code:EXPERT_HELP_NEEDED_TO_CHANGE_ABSTRACTIONS]{Participants
sometimes noted that they would need to ask a domain or visualization expert for
help in order to change data abstractions.} Along with techniques and software
solutions appearing as answers to how the participant would actually transform
the data abstraction, some participants acknowledged they either needed more
information from a data theorist (e.g. ``Could be displayed as a tree, I would
hire someone'') or from a domain expert (``...need to discuss this in more
detail with a domain expert...this data was not provided'').

\codeLabel[code:MORE_DATA_NEEDED_TO_CHANGE_ABSTRACTIONS]{Participants sometimes
noted that more information would need to be collected and added to the data
before transitioning to a different abstraction.} To transform their data from
one abstraction to another, participants stated that they would need to collect
additional data, such as images, speech transcripts, recordings, and labels.

\codeLabel[code:WRANGLING_TOOLS_VERY_DIVERSE]{There was a wide distribution of the tools and techniques that data workers would use to wrangle data.} Survey participants reported 54 different tools by name, with many tools being unique to a single participant. Tools that were mentioned by multiple participants tended to be programming languages.

\noindent\rule[1em]{\linewidth}{0.4pt}  

\vspace{-1em}


\noindent Codes \textbf{C7--C12} are based on evidence from multiple sources,
and are suggestive of unspoken perspectives, intuitions, and fears that may be
common among data workers.

\codeLabel[code:LATENT_WRANGLING_OR_ABSTRACTION]{Even before the survey guided
participants to alternative abstractions, they discussed how they could see
their data in other forms}. This manifested both in conversations with
participants before they took the survey, as well as in comments in the earliest
sections of the survey before the question was asked.

\codeLabel[code:DOESNT_COUNT_AS_DATA]{Many data workers did not feel that what
they work with ``counts as data.''} This comment was a common refrain while
soliciting survey participation at both technical conferences, as well as
through deployment across the university. However, outside of the survey, three
informants (I1, I2, I4) independently made this observation while
reflecting on their experiences working with people new to Data Science. For
example, I2 often runs a data science workshop in the humanities but it tends
to get very low attendance---often the same three participants. Seeing
information as ``data'' may take a certain level of creativity and willingness
to experiment and fail. One Supercomputing survey participant working on
hardware design felt that treating circuit diagrams as ``data'' would be very
strange, and perhaps inappropriate.

\codeLabel[code:SCOPE_CREEP_FEARS]{Thinking about alternative data abstractions
can provoke fears of scope creep.} During a discussion with informants
I6--I12, there was a consensus that exploring alternative abstractions
can be very beneficial for the success of a project, however, it was also
cautioned that it would have the potential to cause misalignments in the vision
of a collaboration---usually termed ``scope creep.'' Data workers are often
cognizant of the impacts that changes to the design of their abstraction will
have, including considerations and costs that they may or may not be able to
articulate in detail.

\codeLabel[code:EMOTIONS_ABOUT_DATA_ABSTRACTIONS]{Data abstractions are often
personal in nature to a data worker.} Based on prior experiences, such as
designing a visualization with I6--I12, the authors recognized that abstractions
can be personal, subjective, and contextual. Wrapped in an existing data
abstraction are a data worker's personal preferences, prior data science
knowledge, and domain knowledge. Thus, suggestions to change this abstraction
are often met with feelings of confusion and resistance. Some of these emotions
stem from concerns about additional work overhead, such as those identified
by~\ref{code:SCOPE_CREEP_FEARS}. Other times, these emotions stem from the
ecosystem of how the data was created, the people it may impact, and the
subjects of the data---all things that a data worker may understand but a
theorist may be unaware of.

\codeLabel[code:INTUITION_ABOUT_NETWORKS]{Data workers often have ``gut
feelings'' or intuition about their data as networks}. Data workers, regardless
of whether their data is known to be network data or not, tended to have some
intuition about the existence of networks within their data, even if specifics such as the meaning of a node or edge were unknown. Special types of networks, such as DAGs and trees, were also mentioned.

\codeLabel[code:INTUITION_ABOUT_GROUPS]{Data workers often have ``gut feelings''
or intuition about their data as clusters, sets, or groups}. Similar
to~\ref{code:INTUITION_ABOUT_NETWORKS}, data workers also had intuition about
the existence of groups in their data. They sometimes referred to hierarchies
existing in and among these groups, and also intuited patterns and clusters in
their data.



\vspace{1em}
\hrule
\vspace{1ex}

\noindent Codes \textbf{C13--C18} highlight informative weaknesses of our typology.

\codeLabel[code:VARIATION_IN_HIERARCHY_PERSPECTIVES]{There is wide variation in
how data workers describe hierarchies.} There was some initial difficulty designing the survey when deciding where hierarchies should fall. Even among the authors, we recognized that one could describe hierarchies as spatial, as networks, as nested sets. We questioned whether a tree and a hierarchy are the same thing, but concluded they have semantic differences. In the final survey, hierarchies were grouped with networks as a ``Network/Hierarchy" abstraction type, with ``Hierarchy'' chosen deliberately to seek feedback. This diversity of perspectives was confirmed; one participant commented that they more closely align hierarchies with groups: ``I find the separation of hierarchies and groupings to be a bit problematic for this domain. Many codes, such as diagnosis codes, exist in a hierarchy (defined by metadata). However it is quite common to refer to areas of this hierarchy as groupings.''

\codeLabel[code:DATASET_A_COMPLEX_AND_EVOLVING_FORM]{Most datasets did not fit
in one category, and participants talked about not just the raw data, but
derived values, metadata, or even ``multiple datasets.''} Participants often
selected multiple data abstractions in response to the initial question of
categorizing their dataset. Heterogeneous datasets are very common, such as when
metadata takes a different form from the main dataset, or when one dataset is a
nested ``value'' inside another of a different type.

\codeLabel[code:MEDIA_TOO_UNDEFINED]{``Media'' as a category had a less
well-defined mental model, resulting in a space with too little structure for
participants to map their data crisply when forced to think of their data as
``media.''} When asked to consider media as an alternative abstraction, a common
response was to imagine screen-capturing to record images and video of a
visualization of the data. But thinking of their data in this way elicited
feelings of discomfort from some participants; comments such as: ``This is weird.
I think of the data not as media but I'm actively trying to turn it into media"
and "I have displayed this data by mapping some of it [to color channels in a
heatmap], but I don't consider the data itself to `be' media or `have' media."
Some data workers understand some sort of inherent visual quality of their data.
For example, one response was ``The data set itself does not include any media,
but interpretations of it are visual in nature... The data could be illustrated
by addition of multidimensional images or 3D meshes when interlinked with
concepts in the graph."

\codeLabel[code:FOREIGN_TERMINOLOGY]{Even very technical data workers find some
data abstraction concepts, language foreign.} We noticed confusion and
misunderstanding surrounding our abstraction terminology; notably, terminology
surrounding tabular data (e.g. items, attributes) was unknown to one
Supercomputing participant and needed to be related to the physical spreadsheet
(e.g. rows, columns) to clarify. This difference in theory-based thinking and
practice-based thinking shows that there is a disconnect between how
visualization people talk about data, and how data workers in general talk about
data.

\codeLabel[code:MISSING_FUNCTION_CATEGORY]{Many data workers consider functions
to be data.} One unexpected finding, after reviewing responses aligning
with~\ref{code:DOESNT_COUNT_AS_DATA}, was that a subset of participants recognize functions as data. These datasets include continuous models, functions
like regression models from housing data, collections of partial differential
equations, or constraint data for linear or integer programming, which I5 and
one author did not consider to be ``spatial'' as defined in the survey.

\codeLabel[code:MISSING_CODE_CATEGORY]{Many data workers consider code to be
data.} As part of a larger discussion about open science and data sharing,
several informants noted that code should be considered data. At a minimum, code
acts as ``metadata'' by providing provenance of where a given dataset came from.
As I6 noted that, ``one person's metadata is another person's data.''

\vspace{1em}
\hrule
\vspace{1ex}

\noindent Codes \textbf{C19--C21} describe the different ways that it was
difficult to focus conversations with data workers on the design of a data
abstraction.

\codeLabel[code:HARD_TO_ISOLATE_DESIGN_FROM_FILE_FORMATS]{The design of a data
abstraction proved difficult to talk about in isolation from specific file
formats.} Related to~\ref{code:FOREIGN_TERMINOLOGY}, some survey participants
misunderstood the connection between an abstraction and its implementation (e.g.
a table vs. a spreadsheet). As a result, in response to our request for ``Other
Generalizations,'' they suggested file formats that were clear fits for our
existing six abstractions such as: ``directed graph represented in a format such
as dot'' instead of Network/Hierarchy, ``CSV file" instead of Tabular, ``a
collection of free text'' instead of Textual.

\codeLabel[code:HARD_TO_ISOLATE_DESIGN_FROM_SOFTWARE_ABSTRACTIONS]{The design of
data abstraction proved difficult to talk about in isolation from software and
programming language abstractions.} One author noted difficulties in focusing
conversations on how a person thinks about their data; informants frequently
pivoted to talking about abstractions imposed by software that were often only
loosely associated with the data model itself, such as git's model of remotes
and branches, or Jupyter's statefulness.

\codeLabel[code:HARD_TO_ISOLATE_DESIGN_FROM_OTHER_CONCERNS]{The design of a data
abstraction proved difficult to talk about in isolation from discovery, capture,
curation, and creation.}~\cite{muller_CuriosityCreativity_2014} Discussions
often detoured from data design to topics such as data provenance and other data
wrangling concerns. Similarly, when prompted to transform their data from one
abstraction to another, some participants suggested collecting entirely new
datasets, rather than transforming the existing data.

\vspace{1em}
\hrule
\vspace{1ex}

\noindent Codes \textbf{C22--C24} describe things that appeared to aid
reflection and communication about data abstractions.

\codeLabel[code:SHOWING_RAW_DATA_HELPS_COMMUNICATION]{Showing real data, such as
a spreadsheet, helps data workers and theorists communicate effectively about
data abstractions.} Many different interactions at community meetups, such as
with I3 and I19, were enhanced by the culture of bringing laptops to show
data and inspect it together.

\codeLabel[code:TYPOLOGIES_AID_REFLECTION]{Data abstraction typologies help data
workers discover latent data abstractions.} Asking questions about a data
abstraction and how it fit, or did not fit, into a typology helped expand data
workers' view of their dataset. One participant noted: ``The questions made me
think more about `the nature' of this dataset. I had always considered it to be
`just tabular' but I realize that there is a hierarchy and geographic data (and
a geographic hierarchy) which I hadn't really considered before. As I type this,
we could layer in time and sets when considering multiple elections.'' Data abstraction typologies can help data workers discover underlying latent abstractions, like hierarchies, or how visualizing their data with additional data abstractions may augment understanding, like adding images to patient records.

\codeLabel[code:TYPOLOGIES_HELP_COMMUNICATION]{Data abstraction typologies help data workers communicate at a sufficient level of detail to design a visualization system.}
We observed this directly with I6--I12. A survey participant also noted that the mental exercise of the survey ``prodded me into thinking about my annotations as more of a central player in the overall visualization as opposed to a secondary thought or supporting contextual element.''
Discussing abstraction typologies helps create a common data design language and reinforces the value that both sides (the data worker and visualization designer) bring to the data problem.

\section{Themes}
\label{sec:theory}

Together, these codes form four overarching themes,
including the prevalence of latent data abstractions, interventionist
impacts that pursuing latent abstractions can have, why many data workers may
express hesitancy to pursue latent abstractions, and benefits that
transparency about data typologies can have for the latent abstraction discovery
process. Here, we enumerate evidence that supports each theme.

\textbf{\textcolor{cbTeaserTeal}{T1}: Latent data abstractions are very common.}
At least initially, raw data formats are not designed in such a way as to
anticipate all abstractions that may be needed or useful, yet even though
abstractions may
not be fully actualized in a computer, data workers are often aware of
meaningful, useful abstractions that they can communicate about without specific
prompts~\ref{code:LATENT_WRANGLING_OR_ABSTRACTION}. Some of these abstractions,
particularly networks~\ref{code:INTUITION_ABOUT_NETWORKS} and
groups~\ref{code:INTUITION_ABOUT_GROUPS}, are intuitive to many data
workers.

This theme validates a
known~\cite{sedlmair_DesignStudy_2012,bigelow_ReflectionsHow_2014} phenomenon
that data rarely has a ``correct'' abstraction, even where predominant file
formats exist; we observed that discrepancies between raw file formats and the
way that a data worker thinks are
common~\ref{code:DISCREPANCY_BETWEEN_MENTAL_MODELS_AND_RAW_DATA}. Instead, data
abstractions have a complex and evolving
form~\ref{code:DATASET_A_COMPLEX_AND_EVOLVING_FORM} that must be explicitly
designed.

The designed nature of data abstractions makes it important to note that neither
data workers nor theorists possess comprehensive knowledge of
all possible latent abstractions, and open-minded communication is necessary for
meaningful, useful abstractions to be discovered. This is true for both parties:
theorists are often aware of abstractions that data workers
might not consider to ``count'' as data~\ref{code:DOESNT_COUNT_AS_DATA}.
Similarly, data workers may be aware of abstractions that theorists do not consider to ``count'' as
data~\ref{code:MISSING_FUNCTION_CATEGORY}~\ref{code:MISSING_CODE_CATEGORY}. Data
workers and theorists may also think about the details of the
same abstraction differently~\ref{code:VARIATION_IN_HIERARCHY_PERSPECTIVES}.
Introducing a typology of data abstractions can expose abstractions that neither
party has considered, in that a typology can contain new abstractions that
data workers may not be aware of, or they may lack new abstractions that
theorists have not
considered~\ref{code:TYPOLOGIES_AID_REFLECTION}.

\textbf{\textcolor{cbTeaserOrange}{T2}: The visualization community identifies
data abstractions for its own transferability needs, but the process of
identifying an abstraction is an intervention with far-reaching effects.}
Collaborations with data workers beyond the visualization research community
stand to benefit---and can be harmed---by the way that both parties introduce,
articulate, and explore data abstractions.

Our data validates that visualization researchers, as theorists, are not
operating in a vacuum; some abstractions that are common in the research
community are intuitive to many data
workers~\ref{code:INTUITION_ABOUT_NETWORKS}~\ref{code:INTUITION_ABOUT_GROUPS}.

However, although these commonalities may be good news for the validity of the
work that visualization researchers perform, there are also areas in which the
culture of visualization research clashes with data workers at large: there is a
often a disconnect between what theorists consider to be data and what data
workers consider to be
data~\ref{code:DOESNT_COUNT_AS_DATA}~\ref{code:MISSING_FUNCTION_CATEGORY}
~\ref{code:MISSING_CODE_CATEGORY}. Disconnects also occur between the language
that theorists use to describe data, and the language that data workers
use~\ref{code:FOREIGN_TERMINOLOGY}. These differences in culture risk
miscommunication at best, but also may risk the development of a
bad collaboration, where either the theorists' goals or the data workers' goals become subordinate.

Consequently, for better or worse, introducing a theoretical perspective is almost always an intervention, and the effects of such interventions can be profound. Because the design of data abstractions is so inextricably linked to the other concerns of data discovery, capture, curation, and creation~\ref{code:HARD_TO_ISOLATE_DESIGN_FROM_OTHER_CONCERNS}, changes to the design of a dataset can result in changes to all of its other aspects. Similarly, influencing a data worker's mental model of their data can have far-reaching practical effects, including disruptions in workflows and changes to the file formats~\ref{code:HARD_TO_ISOLATE_DESIGN_FROM_FILE_FORMATS} and software~\ref{code:HARD_TO_ISOLATE_DESIGN_FROM_SOFTWARE_ABSTRACTIONS} that data workers use.

Data workers are often cognizant of the impacts that changes to the design of their abstraction will have~\ref{code:SCOPE_CREEP_FEARS}, even if they may not be able to fully articulate these impacts in detail~\ref{code:EMOTIONS_ABOUT_DATA_ABSTRACTIONS}.

This is why we predict that \textbf{\textcolor{cbTeaserPurple}{T3}: data workers are less willing to pursue latent data abstractions when the design of an existing abstraction is already fundamental to their workflow.} When there exists a direct mapping between familiar software and the raw data format, efforts to introduce new abstractions will likely meet resistance.

The costs of a changed data abstraction design can include a need to learn new
file formats~\ref{code:HARD_TO_ISOLATE_DESIGN_FROM_FILE_FORMATS} and new
software~\ref{code:HARD_TO_ISOLATE_DESIGN_FROM_SOFTWARE_ABSTRACTIONS} that may
come with the need to learn new software skills such as programming. The tight
coupling between data abstractions, workflows, and software can be seen in the
bespoke wrangling software needs that arise from the combinatoric expansion of
diverse abstractions, diverse workflows, and diverse dataset
scales~\ref{code:WRANGLING_TOOLS_VERY_DIVERSE}~\ref{code:RESPONSE_SIZES}.
However, the added cost is reduced when software practices have not yet been established and investments in learning new skills have not been made. This cost can also be mitigated when theorists are willing and able to provide expert help~\ref{code:EXPERT_HELP_NEEDED_TO_CHANGE_ABSTRACTIONS}, such as wrangling the data to its needed forms.

Similarly, the costs of pursuing latent data abstractions can propagate to other data concerns~\ref{code:HARD_TO_ISOLATE_DESIGN_FROM_OTHER_CONCERNS}, such as the need to collect additional data~\ref{code:MORE_DATA_NEEDED_TO_CHANGE_ABSTRACTIONS}. The fears that data workers often feel~\ref{code:EMOTIONS_ABOUT_DATA_ABSTRACTIONS} and voice~\ref{code:SCOPE_CREEP_FEARS} are suggestive that data abstraction changes can spill over into task abstraction changes that may begin to depart from data workers' actual needs. This potential cost can be an opportunity if care is taken to solicit critique whenever theoretical perspectives are introduced. Such introductions often encourage data workers to provide detailed information about their mental models that they might not otherwise articulate.

Theorists need not wait for such impositions, however, to solicit this kind of targeted feedback. \textbf{\textcolor{cbTeaserPink}{T4}: Like access to real data, introducing a data abstraction typology helps to focus reflection and communication about data abstractions at a level of detail that includes actionable information.}

Our data~\ref{code:SHOWING_RAW_DATA_HELPS_COMMUNICATION} validates the known pitfall~\cite{sedlmair_DesignStudy_2012} in which the lack of access to real data can doom a design study collaboration, because visualization researchers are less likely to have enough actionable information to articulate an accurate data abstraction. It also validates that a culture of data review~\cite{williams_VisualizingMoving_2020}, that is careful to emphasize good communication and transparency about the data abstraction, can compensate for a lack of access to real data because the detailed abstraction is a joint objective that all parties have a stake in.

When theorists take the time to be transparent about their agenda, including the typology that they are attempting to fit a worker's data into, revealing the typology can have similar benefits in that it helps a data worker understand what a theorist is looking for~\ref{code:TYPOLOGIES_HELP_COMMUNICATION}. Introducing typologies can expose data workers to latent abstractions that they may not have considered~\ref{code:TYPOLOGIES_AID_REFLECTION}, and provides an opportunity to provide detailed feedback that might otherwise be left unspoken~\ref{code:MISSING_FUNCTION_CATEGORY}~\ref{code:MISSING_CODE_CATEGORY}~\ref{code:VARIATION_IN_HIERARCHY_PERSPECTIVES}. For example, introducing a typology that is a poor fit in how it subdivides data abstraction categories can serve as an aid to communication, in that it can highlight the detailed ways that a worker considers their data to fit or partially fit more than one abstraction category~\ref{code:DATASET_A_COMPLEX_AND_EVOLVING_FORM}.

Not all shortcomings of a typology are equally beneficial, however. Data abstraction categories that are too general~\ref{code:MEDIA_TOO_UNDEFINED} or rely too heavily upon jargon~\ref{code:FOREIGN_TERMINOLOGY} may have limited utility. These limits are highly contextual, for example, a typology that differentiates between partitions of an abstract mathematical space and regions of a physical three-dimensional space might be useful for a data worker with a rich mathematical background to reflect upon; however, for a worker with less mathematical training, the amount of unfamiliar jargon introduced could inhibit detailed feedback. When introducing a data abstraction typology as an explicit design activity~\cite{mckenna_DesignActivity_2014a}, care should be taken to choose a typology with an appropriate level of granularity and enough accessible concepts to encourage feedback and critique.

\section{Discussion}
\label{sec:discussion}
The codes and themes that we present describe phenomena that are suggestive of guidelines for theorizing about data abstractions. Additionally, it has implications for reporting data abstractions in many kinds of visualization research. We also reflect on our experiences and their implications for the design of data abstraction typologies, and lessons learned from our innovations in survey design and deployment.

\subsection{Guidelines for Pursuing (Latent) Data Abstractions}
\label{sec:guidelines}

Reflecting on the presence of latent data abstractions \textbf{\textcolor{cbTeaserTeal}{(T1)}}, the interventionist nature of defining data abstractions \textbf{\textcolor{cbTeaserOrange}{(T2)}} and in some cases the resistance to it \textbf{\textcolor{cbTeaserPurple}{(T3)}}, and the focusing power of typologies \textbf{\textcolor{cbTeaserPink}{(T4)}}, along with our coded findings in \autoref{sec:codes}, we proffer the following guidelines:

\vspace{1ex}

\noindent\textbf{Data owners and abstraction theorists should collaboratively
probe raw data.} A typical design workflow may have data owners {\em describe}
their data synchronously and then {\em give} one or more data files to the
abstraction theorists for later review. There are several surfaces of loss in
this approach, in which latent information remains latent. Data owners may
forget to review elements of their data. Abstraction theorists may make
assumptions given the data file that are only revisited much later, if at all.
Instead, we recommend that initial meetings with data owners involve the
presentation and collaborative probing of at least one raw dataset.

\vspace{1ex}

\noindent\textbf{Abstraction theorists should introduce the typology and process
that they follow.} Just as theorists can feel lost without exposure to the raw
data, data workers can feel lost when theorists attempt to fit a worker's
project into an opaque typology or framework. For example, if a worker does not
understand, at least at a basic level, that a theorist is attempting to identify
relevant data abstractions before considering visual encodings, workers are
forced to second-guess the theorist's needs. In such a situation, discussing
their data in terms of potential visualization designs may appear to be
\emph{helpful}. As theorists request that workers provide at least one raw
dataset, theorists should also reciprocate by preparing and presenting
sufficient background about what they are hoping to learn or observe.

\vspace{1ex}

\noindent\textbf{Create artifacts that document and convey abstraction details
and demonstrate possible permutations.} We discovered that even in discussion
among the authors, people who had a close working relationship and were operating
from the same typology, there were times when we believed we were discussing the
same abstraction of the data, only to discover we had completely different
assumptions once drawings or classifications were made explicit. Explicitly
stating ideas serves as not only a communication aid, but also as a
method to explore the creative space of possible abstractions and as
documentation for resulting abstractions. Furthermore, writing or drawing such
low-level details can be an effective strategy to ground a derailed conversation
and refocus it back on the design of the data.

\vspace{1ex}

\noindent\textbf{Challenges are an effective means of probing. They require
an artifact to be challenged.} Throughout our interactions with data workers,
we observed that suggesting a concrete abstraction, particularly one that was unlike
how the data worker usually conceptualized their data, elicited rich feedback
about their data and their thinking on it. Responses beginning with phrases
like ``That wouldn't work because...'' or ``That makes no sense'' were
precursors to valuable reflections on their data. Setting up such a response
requires some form of artifact, verbal, pictorial, textual, or otherwise to be
challenged. We recommend such situations be approached sincerely as an honest,
creative exercise towards considering other forms.

\vspace{1ex}

\noindent\textbf{Typologies can serve as a guide to elicit latent elements of
the data abstraction from data workers.} A given typology may not fit all
elements of a particular problem and dataset. However, it provides a corpus of
possible abstractions with which to consider the data. These possibilities can
serve as a jumping point to discuss and challenge possible abstractions of the
data. Through our survey and interviews, we observed that discussions of
fitting the data to various forms evoked more detail about the data itself as
well as provided structure to exploring possible alternative abstractions.

\vspace{1ex}

\noindent\textbf{Document and share the provenance of datasets.}
It is appropriate that a visualization and
analysis solution operates on a brief period of a dataset's lifecycle and
often only a subset of all possible data available. However, it is beneficial to
document the latent elements of the data beyond that directly used by that
solution. The source of the data, the transformations it has gone through, and
related data all provide context. This context can be used to better
understand how the data worker, who is more familiar with all of these
elements, conceptualizes the dataset.

\vspace{1ex}

\noindent\textbf{Assess opportunities inherent in derailments.}
The space of (possibly latent) data abstractions is vast in comparison to the
minimal data abstraction represented in a visualization project. In following
these guidelines, it can be easy for both theorists and workers to feel that the
discussion has become derailed: workers may begin to discuss other data concerns
such as data discovery, capture, curation, or creation. Workers may also discuss
specific software or even prematurely begin to volunteer visualization encodings
and techniques. Similarly, theorists may appear to be exploring esoteric
concepts that do not have a clear application to a worker's project, and their
explorations may threaten to add unnecessary labor to a worker's workload.

These derailments can be an opportunity to gain insight:
First, discussing the design of a dataset has a tendency to prompt communication of important low-level information---even if seemingly unrelated---that workers would not otherwise bring up. Second, workers may actually be speaking on topic, but using seemingly irrelevant language about formats, software, or visualization as proxies that can be revealing about domain conventions or language, as well as revealing a need for the theorist to be more transparent about what they are looking for. Third, seemingly irrelevant topics may be indicative of a high-level mismatch of objectives, differences in perspective, or other miscommunications that could otherwise go unnoticed.

Actively seeking critique from data workers can help to identify a theorist's own derailments. Once derailments are identified, ascertaining the extent to which any of these three opportunities exist can guide a theorist as to whether, when, and how to re-center the conversation.

\vspace{1ex}

\noindent\textbf{Document objectives and revisit them regularly.}
Collaborators often have different high-level expectations, ideas, agendas, and
sub-goals/tasks. This is complicated by the potential for a latent
abstraction---even considering one hypothetically---to change collaborator
perspectives and goals in ways that may not be communicated immediately. We
recommend documenting the objectives of the project, and revisiting those
objectives, especially when derailments are indicative of high-level mismatches.

\vspace{1ex}

\noindent\textbf{Schedule interventions to revisit data abstractions.} The
above guidelines discuss how to make the latent {\em apparent}, but require
the latent exist in the minds of people or the artifacts (e.g., the raw data)
available. However, over the course of the project, all people involved may
discover new facets of the data or incorrect assumptions previously made.
Sometimes these discoveries lead to immediate intervention, but sometimes they
expand the latent space. We recommend scheduling time to revisit, challenge,
and refine data abstractions, given possible discoveries that are latent.

\subsection{Implications for Reporting Data Abstractions}

Our data suggests that providing the expert help that many data workers need can
make visualization researchers more effective collaborators. Until recently, as
we discuss in \autoref{sec:dataAboutWrangling}, performing, documenting, and
reporting on this kind of work may have been difficult to accomplish by itself,
even though there is a great need for published guidance and experience to inform many different
kinds of visualization research.

We expect performing and reporting on detailed, applied data wrangling work better
equips visualization experts to collaborate effectively. Recent acknowledgements of ``Data Transformation,''~\cite{munzner_ReVISeCommittee_2019} ``Data Abstraction,'' and ``Data Structure''~\cite{lee_BroadeningIntellectual_2019} as potential standalone contribution areas may aid in these efforts. We also suggest that such reports may be able to help ground technique- and systems- focused research in more evidence-based user needs.

\subsection{Implications for Designing Abstraction Typologies}

Our experience in attempting to apply the same data abstraction typology to a diverse array of data workers and datasets revealed wide variability in the extent to which typologies are likely to fit a particular context---both the diversity of datasets and the diversity of data worker expertise and perspectives can risk a poor fit.

Our data shows that this is not necessarily problematic. It demonstrates how typologies can be useful in pursuing latent data abstractions despite---and, in some circumstances, \emph{because} of---their
limitations. In the spirit of the observation that ``all models are wrong but some are
useful''~\cite{box_RobustnessStrategy_1979}, shortcomings of a typology can create opportunities to aid in detailed communication and reflection that might be less likely if the typology were a perfect fit.

This also suggests that typologies may not scale well for purposes beyond the
pursuit of latent data abstractions: typologies must generalize in order to be
tractable and support comparison, however, generalizations fundamentally censor
diverse, individual voices and risk stifling important exceptions and innovative
thinking. Our corpus of survey responses demonstrates a way that a conversation about the nature of data abstractions can be conducted at scale, in a way that balances the need for generalizability, while giving
priority to individual viewpoints and grounding discussion in the context of
real-world applications. In the way that our survey explicitly sought critique on the typology that we presented, it allowed for enough organization to visualize, compare, and contrast hundreds of viewpoints, while giving wider freedom for participants to engage directly with its implicit theoretical questions.

\subsection{Reflections on Survey Innovations and Deployment}
\label{sec:reflections}
Unlike typical surveys that primarily collect quantitative information for well-defined questions, our main objective in deploying the survey was to probe for blind spots in our own understanding of what data abstractions exist, and how data workers think about them.

Consequently, we sought to create a survey that was as open-ended as possible.
Closed questions are therefore least ideal, as they provide zero opportunities
for a participant to signal to researchers when there is a problem---researchers
have to anticipate every possible response~\cite{reja_OpenendedVs_2003}.

Open-ended, free response questions at least make it possible for participants
to submit critique, but because they're expensive to code and analyze, and because
they introduce more survey fatigue, they often take the form of a single comment
field at the end that are only used as an ``outlet'' for participants, rather
than a prioritized source of data~\cite{geer_OpenEndedQuestions_1991}.

The extent to which participants freely made use of the ability to skip survey
sections suggests that this approach has several benefits. Replacing a whole
section of a survey with a single free response field appears to help mitigate
survey fatigue. The free response field is at least as open-ended as regular
free response questions, and consequently incurs no additional analysis cost.
The act of stepping outside the normal flow of the survey appears to have
encouraged participants to think about the design of the survey itself, and in
some cases, engage at a theoretical level that more closely resembles a forum
than a survey.

In contrast, our interactive glossary did not appear to have garnered as much
attention---this may have been due to its placement outside the flow of
the survey, and/or its position in the corner of the screen.

The survey innovations created opportunities to improve our understanding of
what data abstractions exist, what terminology is actually used by diverse data
workers, to refine the evolving themes, and we expect it will inform future
iterations of the survey.

\section{Limitations and Future Work}
Here we document the limitations of the survey that we present, our intent to deploy it to a broader audience, and suggest future uses for the dataset that we have released.

\subsection{Survey Design and Evaluation}
The archive of survey responses that we present is not without typical technical difficulties. One major drawback of its design was that the length of the survey varied, depending on the difference between ``rarely'' thinking about a dataset as a certain type and ``never.'' This resulted in some participants filling out lengthier surveys, who showed signs of fatigue. Additionally, a question in several of the Details sections had a bug that failed to capture data completely. Finally, as participants almost always took the survey on their own devices, connectivity and browser incompatibility issues arose, especially for specific iOS devices. These challenges, together with a small number of responses in which participants appeared to abuse the ability to skip, etc., resulted in a set of questionable responses that we flagged.

Rather than suppress these errata, they are included in the archive, and documented in context in the visualized summary of each question. The set of questionable responses can also be interactively filtered out.

As the survey design itself is not our primary contribution, we have only evaluated the extent to which our innovations were effective in achieving our qualitative aims. We can not speak to whether they are effective ways to solicit critique in general, nor engaging enough to encourage theoretical reflection at the levels that we observed.

\subsection{Further Survey Deployment}
The feedback that we collected may also have been influenced by the groups
where we deployed the survey and wrote memos about our observations. The
populations we engaged with during this study all had a high interest in
computing: domain scientists who come to hacking-oriented meetups and attendees
at computing conferences. Although the Supercomputing conference has thousands
of attendees who are there for reasons other than the technical program, in some
interactions, we had difficulty convincing those people that their data counts as
``data.'' Thus, our data and subsequent findings are lacking representation
among people who do not identify with data.

Effectively engaging people with less overt
interest~\cite{peck_DataPersonal_2019}, that may not share the goals represented
by our ``data worker'' persona, is an ongoing effort that we hope to pursue in
future work. Subsequent survey deployment and memo writing will target more
diverse data perspectives and skill sets, by networking with people from
non-Computer Science backgrounds. For example, Meetup attendees have already
referenced ongoing discussions about data abstractions in a paleontology
community. They are considering how to best match and connect competing
ontologies from different sources. Similarly, we have been connected with a
group of vehicle mechanics that are adapting their tables of diagnostic metrics
to changes introduced by increasing numbers of electric vehicles. Other
potential domains include linguistics, sociology, bioinformatics, construction
equipment, and athletics. We intend to advertise and deploy
our survey to more diverse groups of data workers, through academic and
professional conferences, at relevant community Meetup events, and through word
of mouth.

\subsection{Data Reuse}
We have released the public portions of the survey data in a visual, searchable
format as a standalone research contribution, so that individual voices can be
heard and reviewed by researchers studying similar phenomena, beyond our
research aims. Such aims might include creating terminology maps across domains,
using evidence in our survey responses to motivate and justify the design of
general-purpose visualization and data wrangling tools, and other analyses.

\section{Conclusion}
Our grounded theory investigation into the malleability of data abstractions has
resulted in themes that describe data abstractions and their implications for
visualization design, guidelines for the development of data abstractions, the
design and deployment of an open survey, and a corpus of survey responses that
represent a discussion about the nature of data abstractions at scale. This work
has implications for how data abstractions are reported, how typologies are
designed and discussed, and may inform future surveys that seek critique.
Ultimately, this work sheds light on why thinking and communicating about data
abstractions can be difficult, and shows how to best take advantage of
opportunities inherent in that process, as well as mitigate its risks.

\acknowledgments{
The authors wish to thank all participants; informants; Arizona Research Bazaar;
Arizona Research Computing; the Humans, Data, and Computers Lab; and Joshua
Levine. This work was supported by the United States Department of Defense
through DTIC Contract FA8075-14-D-0002-0007, the National Science Foundation
under NSF III-1656958 and NSF III-1844573, and UA Health Sciences through the
Data Science Fellows program.}

\bibliographystyle{abbrv-doi-hyperref-narrow}

\bibliography{main}

\begin{thebibliography}{10}
\renewcommand*{\sfdefault}{PTSansNarrow-TLF}

\bibitem{beyer_ContextualDesign_1999}
\href{https://doi.org/10.1145/291224.291229}{H.~Beyer and K.~Holtzblatt}.
\newblock \href{https://doi.org/10.1145/291224.291229}{Contextual design}.
\newblock \href{https://doi.org/10.1145/291224.291229}{{\em Interactions}},
  \href{https://doi.org/10.1145/291224.291229}{6(1):32--42},
  \href{https://doi.org/10.1145/291224.291229}{Jan. 1999}.
  \href{https://doi.org/10.1145/291224.291229}
{doi: \textsf{%
10\hspace{.1pt}\discretionary{.}{%
}{.}\hspace{.4pt}1145\discretionary{/}{%
}{/}291224\hspace{.1pt}\discretionary{.}{%
}{.}\hspace{.4pt}291229}}


\bibitem{bigelow_DrivingGenetics_2012}
A.~Bigelow.
\newblock {\em Driving Genetics with Experimental Visualization}.
\newblock Undergraduate thesis, University of Utah, Salt Lake City, UT, USA,
  2012.

\bibitem{bigelow_ReflectionsHow_2014}
\href{https://doi.org/10.1145/2598153.2598175}{A.~Bigelow, S.~Drucker,
  D.~Fisher, and M.~Meyer}.
\newblock \href{https://doi.org/10.1145/2598153.2598175}{Reflections on how
  designers design with data}.
\newblock \href{https://doi.org/10.1145/2598153.2598175}{In {\em Proceedings of
  the 2014 International Working Conference on Advanced Visual Interfaces}},
  \href{https://doi.org/10.1145/2598153.2598175}{AVI '14},
  \href{https://doi.org/10.1145/2598153.2598175}{pp. 17--24}.
  \href{https://doi.org/10.1145/2598153.2598175}{ACM},
  \href{https://doi.org/10.1145/2598153.2598175}{2014}.
  \href{https://doi.org/10.1145/2598153.2598175}
{doi: \textsf{%
10\hspace{.1pt}\discretionary{.}{%
}{.}\hspace{.4pt}1145\discretionary{/}{%
}{/}2598153\hspace{.1pt}\discretionary{.}{%
}{.}\hspace{.4pt}2598175}}


\bibitem{bigelow_OrigraphInteractive_2019}
A.~Bigelow, C.~Nobre, M.~D. Meyer, and A.~Lex.
\newblock Origraph: Interactive network wrangling.
\newblock {\em IEEE VAST}, 2019.

\bibitem{box_RobustnessStrategy_1979}
\href{https://doi.org/10.1016/B978-0-12-438150-6.50018-2}{G.~E. Box}.
\newblock \href{https://doi.org/10.1016/B978-0-12-438150-6.50018-2}{Robustness
  in the strategy of scientific model building}.
\newblock \href{https://doi.org/10.1016/B978-0-12-438150-6.50018-2}{In {\em
  Robustness in Statistics}},
  \href{https://doi.org/10.1016/B978-0-12-438150-6.50018-2}{pp. 201--236}.
  \href{https://doi.org/10.1016/B978-0-12-438150-6.50018-2}{Academic Press},
  \href{https://doi.org/10.1016/B978-0-12-438150-6.50018-2}{1979}.
  \href{https://doi.org/10.1016/B978-0-12-438150-6.50018-2}
{doi: \textsf{%
10\hspace{.1pt}\discretionary{.}{%
}{.}\hspace{.4pt}1016\discretionary{/}{%
}{/}B978\discretionary{%
}{-}{-}0\discretionary{%
}{-}{-}12\discretionary{%
}{-}{-}438150\discretionary{%
}{-}{-}6\hspace{.1pt}\discretionary{.}{%
}{.}\hspace{.4pt}50018\discretionary{%
}{-}{-}2}}


\bibitem{carcary_ResearchAudit_2009}
M.~Carcary.
\newblock The research audit trial--enhancing trustworthiness in qualitative
  inquiry.
\newblock {\em Electronic Journal of Business Research Methods}, 7(1), 2009.

\bibitem{charmaz_ConstructingGrounded_2014}
K.~Charmaz.
\newblock {\em Constructing Grounded Theory}.
\newblock sage, 2014.

\bibitem{chi_TaxonomyVisualization_2000}
\href{https://doi.org/10.1109/INFVIS.2000.885092}{E.~Chi}.
\newblock \href{https://doi.org/10.1109/INFVIS.2000.885092}{A taxonomy of
  visualization techniques using the data state reference model}.
\newblock \href{https://doi.org/10.1109/INFVIS.2000.885092}{In {\em IEEE
  Symposium on Information Visualization 2000. INFOVIS 2000. Proceedings}},
  \href{https://doi.org/10.1109/INFVIS.2000.885092}{pp. 69--75},
  \href{https://doi.org/10.1109/INFVIS.2000.885092}{Oct. 2000}.
  \href{https://doi.org/10.1109/INFVIS.2000.885092}
{doi: \textsf{%
10\hspace{.1pt}\discretionary{.}{%
}{.}\hspace{.4pt}1109\discretionary{/}{%
}{/}INFVIS\hspace{.1pt}\discretionary{.}{%
}{.}\hspace{.4pt}2000\hspace{.1pt}\discretionary{.}{%
}{.}\hspace{.4pt}885092}}


\bibitem{dagandra_InformFormInformation_2012}
M.~Da~Gandra and M.~Van~Neck.
\newblock {\em InformForm: Information Design: In Theory, an Informed
  Practice}.
\newblock Mwmcreative Limited, July 2012.

\bibitem{feinberg_DesignPerspective_2017}
\href{https://doi.org/10.1145/3025453.3025837}{M.~Feinberg}.
\newblock \href{https://doi.org/10.1145/3025453.3025837}{A design perspective
  on data}.
\newblock \href{https://doi.org/10.1145/3025453.3025837}{In {\em Proceedings of
  the 2017 CHI Conference on Human Factors in Computing Systems}},
  \href{https://doi.org/10.1145/3025453.3025837}{CHI '17},
  \href{https://doi.org/10.1145/3025453.3025837}{pp. 2952--2963}.
  \href{https://doi.org/10.1145/3025453.3025837}{Association for Computing
  Machinery}, \href{https://doi.org/10.1145/3025453.3025837}{Denver, Colorado,
  USA}, \href{https://doi.org/10.1145/3025453.3025837}{May 2017}.
  \href{https://doi.org/10.1145/3025453.3025837}
{doi: \textsf{%
10\hspace{.1pt}\discretionary{.}{%
}{.}\hspace{.4pt}1145\discretionary{/}{%
}{/}3025453\hspace{.1pt}\discretionary{.}{%
}{.}\hspace{.4pt}3025837}}


\bibitem{figueiras_TypologyData_2013}
\href{https://doi.org/10.1109/IV.2013.45}{A.~Figueiras}.
\newblock \href{https://doi.org/10.1109/IV.2013.45}{A typology for data
  visualization on the web}.
\newblock \href{https://doi.org/10.1109/IV.2013.45}{In {\em 2013 17th
  International Conference on Information Visualisation}},
  \href{https://doi.org/10.1109/IV.2013.45}{pp. 351--358},
  \href{https://doi.org/10.1109/IV.2013.45}{July 2013}.
  \href{https://doi.org/10.1109/IV.2013.45}
{doi: \textsf{%
10\hspace{.1pt}\discretionary{.}{%
}{.}\hspace{.4pt}1109\discretionary{/}{%
}{/}IV\hspace{.1pt}\discretionary{.}{%
}{.}\hspace{.4pt}2013\hspace{.1pt}\discretionary{.}{%
}{.}\hspace{.4pt}45}}


\bibitem{geer_OpenEndedQuestions_1991}
\href{https://doi.org/10.1086/269268}{J.~G. Geer}.
\newblock \href{https://doi.org/10.1086/269268}{Do open-ended questions measure
  "salient" issues?}
\newblock \href{https://doi.org/10.1086/269268}{{\em Public Opinion
  Quarterly}}, \href{https://doi.org/10.1086/269268}{55(3):360--370},
  \href{https://doi.org/10.1086/269268}{Jan. 1991}.
  \href{https://doi.org/10.1086/269268}
{doi: \textsf{%
10\hspace{.1pt}\discretionary{.}{%
}{.}\hspace{.4pt}1086\discretionary{/}{%
}{/}269268}}


\bibitem{goodwin_CreativeUserCentered_2013}
\href{https://doi.org/10.1109/TVCG.2013.145}{S.~Goodwin, J.~Dykes, S.~Jones,
  I.~Dillingham, G.~Dove, A.~Duffy, A.~Kachkaev, A.~Slingsby, and J.~Wood}.
\newblock \href{https://doi.org/10.1109/TVCG.2013.145}{Creative user-centered
  visualization design for energy analysts and modelers}.
\newblock \href{https://doi.org/10.1109/TVCG.2013.145}{{\em IEEE Transactions
  on Visualization and Computer Graphics}},
  \href{https://doi.org/10.1109/TVCG.2013.145}{19(12):2516--2525},
  \href{https://doi.org/10.1109/TVCG.2013.145}{Dec. 2013}.
  \href{https://doi.org/10.1109/TVCG.2013.145}
{doi: \textsf{%
10\hspace{.1pt}\discretionary{.}{%
}{.}\hspace{.4pt}1109\discretionary{/}{%
}{/}TVCG\hspace{.1pt}\discretionary{.}{%
}{.}\hspace{.4pt}2013\hspace{.1pt}\discretionary{.}{%
}{.}\hspace{.4pt}145}}


\bibitem{guo_ProactiveWrangling_2011}
\href{https://doi.org/10.1145/2047196.2047205}{P.~J. Guo, S.~Kandel, J.~M.
  Hellerstein, and J.~Heer}.
\newblock \href{https://doi.org/10.1145/2047196.2047205}{Proactive wrangling:
  Mixed-initiative end-user programming of data transformation scripts}.
\newblock \href{https://doi.org/10.1145/2047196.2047205}{In {\em Proceedings of
  the 24th Annual ACM Symposium on User Interface Software and Technology}},
  \href{https://doi.org/10.1145/2047196.2047205}{UIST '11},
  \href{https://doi.org/10.1145/2047196.2047205}{pp. 65--74}.
  \href{https://doi.org/10.1145/2047196.2047205}{Association for Computing
  Machinery}, \href{https://doi.org/10.1145/2047196.2047205}{Santa Barbara,
  California, USA}, \href{https://doi.org/10.1145/2047196.2047205}{Oct. 2011}.
  \href{https://doi.org/10.1145/2047196.2047205}
{doi: \textsf{%
10\hspace{.1pt}\discretionary{.}{%
}{.}\hspace{.4pt}1145\discretionary{/}{%
}{/}2047196\hspace{.1pt}\discretionary{.}{%
}{.}\hspace{.4pt}2047205}}


\bibitem{hanington_UniversalMethods_2012}
B.~Hanington and B.~Martin.
\newblock {\em Universal Methods of Design:100 Ways to Research Complex
  Problems, Develop Innovative Ideas, and Design Effective Solutions}.
\newblock Rockport Publishers, Feb. 2012.

\bibitem{heer_OrionSystem_2011a}
J.~Heer and A.~Perer.
\newblock Orion: A system for modeling, transformation and visualization of
  multidimensional heterogeneous networks.
\newblock In {\em IEEE Visual Analytics Science \textbackslash\& Technology
  (VAST)}, p.~10, 2011.

\bibitem{kandel_WranglerInteractive_2011}
\href{https://doi.org/10.1145/1978942.1979444}{S.~Kandel, A.~Paepcke,
  J.~Hellerstein, and J.~Heer}.
\newblock \href{https://doi.org/10.1145/1978942.1979444}{Wrangler: Interactive
  visual specification of data transformation scripts}.
\newblock \href{https://doi.org/10.1145/1978942.1979444}{{\em Proceedings of
  the SIGCHI Conference on Human Factors in Computing Systems}},
  \href{https://doi.org/10.1145/1978942.1979444}{pp. 3363--3372},
  \href{https://doi.org/10.1145/1978942.1979444}{2011}.
  \href{https://doi.org/10.1145/1978942.1979444}
{doi: \textsf{%
10\hspace{.1pt}\discretionary{.}{%
}{.}\hspace{.4pt}1145\discretionary{/}{%
}{/}1978942\hspace{.1pt}\discretionary{.}{%
}{.}\hspace{.4pt}1979444}}


\bibitem{kerzner_FrameworkCreative_2019}
\href{https://doi.org/10.1109/TVCG.2018.2865241}{E.~Kerzner, S.~Goodwin,
  J.~Dykes, S.~Jones, and M.~Meyer}.
\newblock \href{https://doi.org/10.1109/TVCG.2018.2865241}{A framework for
  creative visualization-opportunities workshops}.
\newblock \href{https://doi.org/10.1109/TVCG.2018.2865241}{{\em IEEE
  Transactions on Visualization and Computer Graphics}},
  \href{https://doi.org/10.1109/TVCG.2018.2865241}{25(1):748--758},
  \href{https://doi.org/10.1109/TVCG.2018.2865241}{Jan. 2019}.
  \href{https://doi.org/10.1109/TVCG.2018.2865241}
{doi: \textsf{%
10\hspace{.1pt}\discretionary{.}{%
}{.}\hspace{.4pt}1109\discretionary{/}{%
}{/}TVCG\hspace{.1pt}\discretionary{.}{%
}{.}\hspace{.4pt}2018\hspace{.1pt}\discretionary{.}{%
}{.}\hspace{.4pt}2865241}}


\bibitem{kiss_EvaluationManually_2013}
\href{https://doi.org/10.1145/2501105.2501106}{{\'A}.~Kiss and
  T.~Szir{\'a}nyi}.
\newblock \href{https://doi.org/10.1145/2501105.2501106}{Evaluation of manually
  created ground truth for multi-view people localization}.
\newblock \href{https://doi.org/10.1145/2501105.2501106}{In {\em Proceedings of
  the International Workshop on Video and Image Ground Truth in Computer Vision
  Applications}}, \href{https://doi.org/10.1145/2501105.2501106}{VIGTA '13},
  \href{https://doi.org/10.1145/2501105.2501106}{pp. 1--6}.
  \href{https://doi.org/10.1145/2501105.2501106}{Association for Computing
  Machinery}, \href{https://doi.org/10.1145/2501105.2501106}{St. Petersburg,
  Russia}, \href{https://doi.org/10.1145/2501105.2501106}{July 2013}.
  \href{https://doi.org/10.1145/2501105.2501106}
{doi: \textsf{%
10\hspace{.1pt}\discretionary{.}{%
}{.}\hspace{.4pt}1145\discretionary{/}{%
}{/}2501105\hspace{.1pt}\discretionary{.}{%
}{.}\hspace{.4pt}2501106}}


\bibitem{lee_BroadeningIntellectual_2019}
\href{https://doi.org/10.1109/MCG.2019.2914844}{B.~Lee, K.~Isaacs, D.~A.
  Szafir, G.~E. Marai, C.~Turkay, M.~Tory, S.~Carpendale, and A.~Endert}.
\newblock \href{https://doi.org/10.1109/MCG.2019.2914844}{Broadening
  intellectual diversity in visualization research papers}.
\newblock \href{https://doi.org/10.1109/MCG.2019.2914844}{{\em IEEE Computer
  Graphics and Applications}},
  \href{https://doi.org/10.1109/MCG.2019.2914844}{39(4):78--85},
  \href{https://doi.org/10.1109/MCG.2019.2914844}{July 2019}.
  \href{https://doi.org/10.1109/MCG.2019.2914844}
{doi: \textsf{%
10\hspace{.1pt}\discretionary{.}{%
}{.}\hspace{.4pt}1109\discretionary{/}{%
}{/}MCG\hspace{.1pt}\discretionary{.}{%
}{.}\hspace{.4pt}2019\hspace{.1pt}\discretionary{.}{%
}{.}\hspace{.4pt}2914844}}


\bibitem{liu_PloceusModeling_2014}
\href{https://doi.org/10.1177/1473871613488591}{Z.~Liu, S.~B. Navathe, and
  J.~T. Stasko}.
\newblock \href{https://doi.org/10.1177/1473871613488591}{Ploceus: Modeling,
  visualizing, and analyzing tabular data as networks}.
\newblock \href{https://doi.org/10.1177/1473871613488591}{{\em Information
  Visualization}},
  \href{https://doi.org/10.1177/1473871613488591}{13(1):59--89},
  \href{https://doi.org/10.1177/1473871613488591}{Jan. 2014}.
  \href{https://doi.org/10.1177/1473871613488591}
{doi: \textsf{%
10\hspace{.1pt}\discretionary{.}{%
}{.}\hspace{.4pt}1177\discretionary{/}{%
}{/}1473871613488591}}


\bibitem{mccurdy_ActionDesign_2016a}
\href{https://doi.org/10.1145/2993901.2993916}{N.~McCurdy, J.~Dykes, and
  M.~Meyer}.
\newblock \href{https://doi.org/10.1145/2993901.2993916}{Action design research
  and visualization design}.
\newblock \href{https://doi.org/10.1145/2993901.2993916}{In {\em Proceedings of
  the Sixth Workshop on Beyond Time and Errors on Novel Evaluation Methods for
  Visualization}}, \href{https://doi.org/10.1145/2993901.2993916}{BELIV '16},
  \href{https://doi.org/10.1145/2993901.2993916}{pp. 10--18}.
  \href{https://doi.org/10.1145/2993901.2993916}{Association for Computing
  Machinery}, \href{https://doi.org/10.1145/2993901.2993916}{Baltimore, MD,
  USA}, \href{https://doi.org/10.1145/2993901.2993916}{Oct. 2016}.
  \href{https://doi.org/10.1145/2993901.2993916}
{doi: \textsf{%
10\hspace{.1pt}\discretionary{.}{%
}{.}\hspace{.4pt}1145\discretionary{/}{%
}{/}2993901\hspace{.1pt}\discretionary{.}{%
}{.}\hspace{.4pt}2993916}}


\bibitem{mckenna_DesignActivity_2017a}
S.~McKenna.
\newblock {\em The Design Activity Framework: Investigating the Data
  Visualization Design Process}.
\newblock PhD thesis, University of Utah, June 2017.

\bibitem{mckenna_DesignActivity_2014a}
\href{https://doi.org/10.1109/TVCG.2014.2346331}{S.~McKenna, D.~Mazur,
  J.~Agutter, and M.~Meyer}.
\newblock \href{https://doi.org/10.1109/TVCG.2014.2346331}{Design activity
  framework for visualization design}.
\newblock \href{https://doi.org/10.1109/TVCG.2014.2346331}{{\em IEEE
  Transactions on Visualization and Computer Graphics}},
  \href{https://doi.org/10.1109/TVCG.2014.2346331}{20(12):2191--2200},
  \href{https://doi.org/10.1109/TVCG.2014.2346331}{Dec. 2014}.
  \href{https://doi.org/10.1109/TVCG.2014.2346331}
{doi: \textsf{%
10\hspace{.1pt}\discretionary{.}{%
}{.}\hspace{.4pt}1109\discretionary{/}{%
}{/}TVCG\hspace{.1pt}\discretionary{.}{%
}{.}\hspace{.4pt}2014\hspace{.1pt}\discretionary{.}{%
}{.}\hspace{.4pt}2346331}}


\bibitem{meyer_CriteriaRigor_2019}
\href{https://doi.org/10.1109/TVCG.2019.2934539}{M.~Meyer and J.~Dykes}.
\newblock \href{https://doi.org/10.1109/TVCG.2019.2934539}{Criteria for rigor
  in visualization design study}.
\newblock \href{https://doi.org/10.1109/TVCG.2019.2934539}{{\em IEEE
  Transactions on Visualization and Computer Graphics}},
  \href{https://doi.org/10.1109/TVCG.2019.2934539}{pp. 1--1},
  \href{https://doi.org/10.1109/TVCG.2019.2934539}{2019}.
  \href{https://doi.org/10.1109/TVCG.2019.2934539}
{doi: \textsf{%
10\hspace{.1pt}\discretionary{.}{%
}{.}\hspace{.4pt}1109\discretionary{/}{%
}{/}TVCG\hspace{.1pt}\discretionary{.}{%
}{.}\hspace{.4pt}2019\hspace{.1pt}\discretionary{.}{%
}{.}\hspace{.4pt}2934539}}


\bibitem{meyer_NestedBlocks_2015}
\href{https://doi.org/10.1177/1473871613510429}{M.~Meyer, M.~Sedlmair, P.~S.
  Quinan, and T.~Munzner}.
\newblock \href{https://doi.org/10.1177/1473871613510429}{The nested blocks and
  guidelines model}.
\newblock \href{https://doi.org/10.1177/1473871613510429}{{\em Information
  Visualization}},
  \href{https://doi.org/10.1177/1473871613510429}{14(3):234--249},
  \href{https://doi.org/10.1177/1473871613510429}{July 2015}.
  \href{https://doi.org/10.1177/1473871613510429}
{doi: \textsf{%
10\hspace{.1pt}\discretionary{.}{%
}{.}\hspace{.4pt}1177\discretionary{/}{%
}{/}1473871613510429}}


\bibitem{moravcsik_ActiveCitation_2010}
\href{https://doi.org/10.1017/S1049096510990781}{A.~Moravcsik}.
\newblock \href{https://doi.org/10.1017/S1049096510990781}{Active citation: A
  precondition for replicable qualitative research}.
\newblock \href{https://doi.org/10.1017/S1049096510990781}{{\em PS: Political
  Science \& Politics}},
  \href{https://doi.org/10.1017/S1049096510990781}{43(1):29--35},
  \href{https://doi.org/10.1017/S1049096510990781}{Jan. 2010}.
  \href{https://doi.org/10.1017/S1049096510990781}
{doi: \textsf{%
10\hspace{.1pt}\discretionary{.}{%
}{.}\hspace{.4pt}1017\discretionary{/}{%
}{/}S1049096510990781}}


\bibitem{muller_CuriosityCreativity_2014}
\href{https://doi.org/10.1007/978-1-4939-0378-8_2}{M.~Muller}.
\newblock \href{https://doi.org/10.1007/978-1-4939-0378-8_2}{Curiosity,
  creativity, and surprise as analytic tools: Grounded theory method}.
\newblock \href{https://doi.org/10.1007/978-1-4939-0378-8_2}{In J.~S. Olson and
  W.~A. Kellogg, eds., {\em Ways of Knowing in HCI}},
  \href{https://doi.org/10.1007/978-1-4939-0378-8_2}{pp. 25--48}.
  \href{https://doi.org/10.1007/978-1-4939-0378-8_2}{Springer},
  \href{https://doi.org/10.1007/978-1-4939-0378-8_2}{New York, NY},
  \href{https://doi.org/10.1007/978-1-4939-0378-8_2}{2014}.
  \href{https://doi.org/10.1007/978-1-4939-0378-8_2}
{doi: \textsf{%
10\hspace{.1pt}\discretionary{.}{%
}{.}\hspace{.4pt}1007\discretionary{/}{%
}{/}978\discretionary{%
}{-}{-}1\discretionary{%
}{-}{-}4939\discretionary{%
}{-}{-}0378\discretionary{%
}{-}{-}8\_2}}


\bibitem{muller_HowData_2019}
\href{https://doi.org/10.1145/3290605.3300356}{M.~Muller, I.~Lange, D.~Wang,
  D.~Piorkowski, J.~Tsay, Q.~V. Liao, C.~Dugan, and T.~Erickson}.
\newblock \href{https://doi.org/10.1145/3290605.3300356}{How data science
  workers work with data: Discovery, capture, curation, design, creation}.
\newblock \href{https://doi.org/10.1145/3290605.3300356}{In {\em Proceedings of
  the 2019 CHI Conference on Human Factors in Computing Systems}},
  \href{https://doi.org/10.1145/3290605.3300356}{CHI '19},
  \href{https://doi.org/10.1145/3290605.3300356}{pp. 126:1--126:15}.
  \href{https://doi.org/10.1145/3290605.3300356}{ACM},
  \href{https://doi.org/10.1145/3290605.3300356}{2019}.
  \href{https://doi.org/10.1145/3290605.3300356}
{doi: \textsf{%
10\hspace{.1pt}\discretionary{.}{%
}{.}\hspace{.4pt}1145\discretionary{/}{%
}{/}3290605\hspace{.1pt}\discretionary{.}{%
}{.}\hspace{.4pt}3300356}}


\bibitem{munzner_NestedModel_2009}
\href{https://doi.org/10.1109/TVCG.2009.111}{T.~Munzner}.
\newblock \href{https://doi.org/10.1109/TVCG.2009.111}{A nested model for
  visualization design and validation}.
\newblock \href{https://doi.org/10.1109/TVCG.2009.111}{{\em IEEE Transactions
  on Visualization and Computer Graphics}},
  \href{https://doi.org/10.1109/TVCG.2009.111}{15(6):921--928},
  \href{https://doi.org/10.1109/TVCG.2009.111}{Nov. 2009}.
  \href{https://doi.org/10.1109/TVCG.2009.111}
{doi: \textsf{%
10\hspace{.1pt}\discretionary{.}{%
}{.}\hspace{.4pt}1109\discretionary{/}{%
}{/}TVCG\hspace{.1pt}\discretionary{.}{%
}{.}\hspace{.4pt}2009\hspace{.1pt}\discretionary{.}{%
}{.}\hspace{.4pt}111}}


\bibitem{munzner_WhatData_2014}
T.~Munzner.
\newblock What: Data abstraction.
\newblock In {\em Visualization Analysis and Design}. CRC Press, Dec. 2014.

\bibitem{munzner_ReVISeCommittee_2019}
T.~Munzner, A.~Endert, A.~Lex, A.~Ynnerman, C.~Garth, M.~Chen, P.~Isenberg, and
  L.~Shixia.
\newblock Revise committee town hall.
\newblock
  \url{https://drive.google.com/drive/u/0/folders/1dqssldHbXLmAD9zeOqHCbfNTb8gjeHKS},
  Oct. 2019.

\bibitem{nielsen_ABySSExplorerVisualizing_2009}
\href{https://doi.org/10.1109/TVCG.2009.116}{C.~Nielsen, S.~Jackman, I.~Birol,
  and S.~Jones}.
\newblock \href{https://doi.org/10.1109/TVCG.2009.116}{Abyss-explorer:
  Visualizing genome sequence assemblies}.
\newblock \href{https://doi.org/10.1109/TVCG.2009.116}{{\em IEEE Transactions
  on Visualization and Computer Graphics}},
  \href{https://doi.org/10.1109/TVCG.2009.116}{15(6):881--888},
  \href{https://doi.org/10.1109/TVCG.2009.116}{Nov. 2009}.
  \href{https://doi.org/10.1109/TVCG.2009.116}
{doi: \textsf{%
10\hspace{.1pt}\discretionary{.}{%
}{.}\hspace{.4pt}1109\discretionary{/}{%
}{/}TVCG\hspace{.1pt}\discretionary{.}{%
}{.}\hspace{.4pt}2009\hspace{.1pt}\discretionary{.}{%
}{.}\hspace{.4pt}116}}


\bibitem{passi_DataVision_2017}
\href{https://doi.org/10.1145/2998181.2998331}{S.~Passi and S.~Jackson}.
\newblock \href{https://doi.org/10.1145/2998181.2998331}{Data vision: Learning
  to see through algorithmic abstraction}.
\newblock \href{https://doi.org/10.1145/2998181.2998331}{In {\em Proceedings of
  the 2017 ACM Conference on Computer Supported Cooperative Work and Social
  Computing - CSCW '17}}, \href{https://doi.org/10.1145/2998181.2998331}{pp.
  2436--2447}. \href{https://doi.org/10.1145/2998181.2998331}{ACM Press},
  \href{https://doi.org/10.1145/2998181.2998331}{Portland, Oregon, USA},
  \href{https://doi.org/10.1145/2998181.2998331}{2017}.
  \href{https://doi.org/10.1145/2998181.2998331}
{doi: \textsf{%
10\hspace{.1pt}\discretionary{.}{%
}{.}\hspace{.4pt}1145\discretionary{/}{%
}{/}2998181\hspace{.1pt}\discretionary{.}{%
}{.}\hspace{.4pt}2998331}}


\bibitem{passi_TrustData_2018}
\href{https://doi.org/10.1145/3274405}{S.~Passi and S.~J. Jackson}.
\newblock \href{https://doi.org/10.1145/3274405}{Trust in data science:
  Collaboration, translation, and accountability in corporate data science
  projects}.
\newblock \href{https://doi.org/10.1145/3274405}{{\em Proceedings of the ACM on
  Human-Computer Interaction}},
  \href{https://doi.org/10.1145/3274405}{2(CSCW):136:1--136:28},
  \href{https://doi.org/10.1145/3274405}{Nov. 2018}.
  \href{https://doi.org/10.1145/3274405}
{doi: \textsf{%
10\hspace{.1pt}\discretionary{.}{%
}{.}\hspace{.4pt}1145\discretionary{/}{%
}{/}3274405}}


\bibitem{peck_DataPersonal_2019}
\href{https://doi.org/10.1145/3290605.3300474}{E.~M. Peck, S.~E. Ayuso, and
  O.~{El-Etr}}.
\newblock \href{https://doi.org/10.1145/3290605.3300474}{Data is personal:
  Attitudes and perceptions of data visualization in rural pennsylvania}.
\newblock \href{https://doi.org/10.1145/3290605.3300474}{In {\em Proceedings of
  the 2019 CHI Conference on Human Factors in Computing Systems}},
  \href{https://doi.org/10.1145/3290605.3300474}{CHI '19},
  \href{https://doi.org/10.1145/3290605.3300474}{pp. 1--12}.
  \href{https://doi.org/10.1145/3290605.3300474}{Association for Computing
  Machinery}, \href{https://doi.org/10.1145/3290605.3300474}{Glasgow, Scotland
  Uk}, \href{https://doi.org/10.1145/3290605.3300474}{May 2019}.
  \href{https://doi.org/10.1145/3290605.3300474}
{doi: \textsf{%
10\hspace{.1pt}\discretionary{.}{%
}{.}\hspace{.4pt}1145\discretionary{/}{%
}{/}3290605\hspace{.1pt}\discretionary{.}{%
}{.}\hspace{.4pt}3300474}}


\bibitem{pine_PoliticsMeasurement_2015}
\href{https://doi.org/10.1145/2702123.2702298}{K.~H. Pine and M.~Liboiron}.
\newblock \href{https://doi.org/10.1145/2702123.2702298}{The politics of
  measurement and action}.
\newblock \href{https://doi.org/10.1145/2702123.2702298}{In {\em Proceedings of
  the 33rd Annual ACM Conference on Human Factors in Computing Systems}},
  \href{https://doi.org/10.1145/2702123.2702298}{CHI '15},
  \href{https://doi.org/10.1145/2702123.2702298}{pp. 3147--3156}.
  \href{https://doi.org/10.1145/2702123.2702298}{Association for Computing
  Machinery}, \href{https://doi.org/10.1145/2702123.2702298}{Seoul, Republic of
  Korea}, \href{https://doi.org/10.1145/2702123.2702298}{Apr. 2015}.
  \href{https://doi.org/10.1145/2702123.2702298}
{doi: \textsf{%
10\hspace{.1pt}\discretionary{.}{%
}{.}\hspace{.4pt}1145\discretionary{/}{%
}{/}2702123\hspace{.1pt}\discretionary{.}{%
}{.}\hspace{.4pt}2702298}}


\bibitem{pretorius_WhatDoes_2009}
\href{https://doi.org/10.1057/ivs.2009.13}{A.~J. Pretorius and J.~J. Van~Wijk}.
\newblock \href{https://doi.org/10.1057/ivs.2009.13}{What does the user want to
  see? what do the data want to be?}
\newblock \href{https://doi.org/10.1057/ivs.2009.13}{{\em Information
  Visualization}}, \href{https://doi.org/10.1057/ivs.2009.13}{8(3):153--166},
  \href{https://doi.org/10.1057/ivs.2009.13}{Sept. 2009}.
  \href{https://doi.org/10.1057/ivs.2009.13}
{doi: \textsf{%
10\hspace{.1pt}\discretionary{.}{%
}{.}\hspace{.4pt}1057\discretionary{/}{%
}{/}ivs\hspace{.1pt}\discretionary{.}{%
}{.}\hspace{.4pt}2009\hspace{.1pt}\discretionary{.}{%
}{.}\hspace{.4pt}13}}


\bibitem{reja_OpenendedVs_2003}
U.~Reja, K.~L. Manfreda, V.~Hlebec, and V.~Vehovar.
\newblock Open-ended vs. close-ended questions in web questionnaires.
\newblock {\em Developments in Applied Statistics}, 19(1):159--177, 2003.

\bibitem{sandelowski_RealQualitative_2001}
\href{https://doi.org/10.1002/nur.1025}{M.~Sandelowski}.
\newblock \href{https://doi.org/10.1002/nur.1025}{Real qualitative researchers
  do not count: The use of numbers in qualitative research}.
\newblock \href{https://doi.org/10.1002/nur.1025}{{\em Research in Nursing \&
  Health}}, \href{https://doi.org/10.1002/nur.1025}{24(3):230--240},
  \href{https://doi.org/10.1002/nur.1025}{June 2001}.
  \href{https://doi.org/10.1002/nur.1025}
{doi: \textsf{%
10\hspace{.1pt}\discretionary{.}{%
}{.}\hspace{.4pt}1002\discretionary{/}{%
}{/}nur\hspace{.1pt}\discretionary{.}{%
}{.}\hspace{.4pt}1025}}


\bibitem{sedlmair_DesignStudy_2012}
M.~Sedlmair, M.~Meyer, and T.~Munzner.
\newblock Design study methodology: Reflections from the trenches and the
  statcks.
\newblock {\em IEEE Transactions on Visualization and Computer Graphics},
  18(12):2431--2440, 2012.

\bibitem{srinivasan_GraphitiInteractive_2018a}
\href{https://doi.org/10.1109/TVCG.2017.2744843}{A.~Srinivasan, H.~Park,
  A.~Endert, and R.~C. Basole}.
\newblock \href{https://doi.org/10.1109/TVCG.2017.2744843}{Graphiti:
  Interactive specification of attribute-based edges for network modeling and
  visualization}.
\newblock \href{https://doi.org/10.1109/TVCG.2017.2744843}{{\em IEEE
  Transactions on Visualization and Computer Graphics}},
  \href{https://doi.org/10.1109/TVCG.2017.2744843}{24(1):226--235},
  \href{https://doi.org/10.1109/TVCG.2017.2744843}{Jan. 2018}.
  \href{https://doi.org/10.1109/TVCG.2017.2744843}
{doi: \textsf{%
10\hspace{.1pt}\discretionary{.}{%
}{.}\hspace{.4pt}1109\discretionary{/}{%
}{/}TVCG\hspace{.1pt}\discretionary{.}{%
}{.}\hspace{.4pt}2017\hspace{.1pt}\discretionary{.}{%
}{.}\hspace{.4pt}2744843}}


\bibitem{sternberg_HandbookCreativity_1999}
R.~J. Sternberg.
\newblock {\em Handbook of Creativity}.
\newblock Cambridge University Press, 1999.

\bibitem{verborgh_UsingOpenRefine_2013}
R.~Verborgh and M.~D. Wilde.
\newblock {\em Using OpenRefine}.
\newblock Packt Publishing Ltd, Sept. 2013.

\bibitem{vonoech_KickSeat_1986}
R.~Von~Oech and G.~Willett.
\newblock {\em A Kick in the Seat of the Pants: Using Your Explorer, Artist,
  Judge, \& Warrior to Be More Creative}.
\newblock Perennial Library, 1986.

\bibitem{walny_DataChanges_2019}
\href{https://doi.org/10.1109/TVCG.2019.2934538}{J.~Walny, C.~Frisson, M.~West,
  D.~Kosminsky, S.~Knudsen, S.~Carpendale, and W.~Willett}.
\newblock \href{https://doi.org/10.1109/TVCG.2019.2934538}{Data changes
  everything: Challenges and opportunities in data visualization design
  handoff}.
\newblock \href{https://doi.org/10.1109/TVCG.2019.2934538}{{\em IEEE
  transactions on visualization and computer graphics}},
  \href{https://doi.org/10.1109/TVCG.2019.2934538}{26(1):12--22},
  \href{https://doi.org/10.1109/TVCG.2019.2934538}{2019}.
  \href{https://doi.org/10.1109/TVCG.2019.2934538}
{doi: \textsf{%
10\hspace{.1pt}\discretionary{.}{%
}{.}\hspace{.4pt}1109\discretionary{/}{%
}{/}TVCG\hspace{.1pt}\discretionary{.}{%
}{.}\hspace{.4pt}2019\hspace{.1pt}\discretionary{.}{%
}{.}\hspace{.4pt}2934538}}


\bibitem{weed_CapturingEssence_2017}
\href{https://doi.org/10.1080/2159676X.2016.1251701}{M.~Weed}.
\newblock \href{https://doi.org/10.1080/2159676X.2016.1251701}{Capturing the
  essence of grounded theory: The importance of understanding commonalities and
  variants}.
\newblock \href{https://doi.org/10.1080/2159676X.2016.1251701}{{\em Qualitative
  Research in Sport, Exercise and Health}},
  \href{https://doi.org/10.1080/2159676X.2016.1251701}{9(1):149--156},
  \href{https://doi.org/10.1080/2159676X.2016.1251701}{Jan. 2017}.
  \href{https://doi.org/10.1080/2159676X.2016.1251701}
{doi: \textsf{%
10\hspace{.1pt}\discretionary{.}{%
}{.}\hspace{.4pt}1080\discretionary{/}{%
}{/}2159676X\hspace{.1pt}\discretionary{.}{%
}{.}\hspace{.4pt}2016\hspace{.1pt}\discretionary{.}{%
}{.}\hspace{.4pt}1251701}}


\bibitem{wiener_MakingTeams_2007}
\href{https://doi.org/10.4135/9781848607941.n14}{C.~Wiener}.
\newblock \href{https://doi.org/10.4135/9781848607941.n14}{{\em Making Teams
  Work in Conducting Grounded Theory}},
  \href{https://doi.org/10.4135/9781848607941.n14}{pp. 292--310}.
\newblock \href{https://doi.org/10.4135/9781848607941.n14}{SAGE Publications
  Ltd}, \href{https://doi.org/10.4135/9781848607941.n14}{1 Oliver's Yard,~55
  City Road,~London~England~EC1Y 1SP~United Kingdom},
  \href{https://doi.org/10.4135/9781848607941.n14}{2007}.
  \href{https://doi.org/10.4135/9781848607941.n14}
{doi: \textsf{%
10\hspace{.1pt}\discretionary{.}{%
}{.}\hspace{.4pt}4135\discretionary{/}{%
}{/}9781848607941\hspace{.1pt}\discretionary{.}{%
}{.}\hspace{.4pt}n14}}


\bibitem{williams_VisualizingMoving_2020}
\href{https://doi.org/10.1109/TVCG.2019.2934285}{K.~Williams, A.~Bigelow, and
  K.~E. Isaacs}.
\newblock \href{https://doi.org/10.1109/TVCG.2019.2934285}{Visualizing a moving
  target: A design study on task parallel programs in the presence of evolving
  data and concerns}.
\newblock \href{https://doi.org/10.1109/TVCG.2019.2934285}{{\em To appear in
  IEEE Transactions on Visualization and Computer Graphics (Proceedings of
  InfoVis '19)}}, \href{https://doi.org/10.1109/TVCG.2019.2934285}{Jan. 2020}.
  \href{https://doi.org/10.1109/TVCG.2019.2934285}
{doi: \textsf{%
10\hspace{.1pt}\discretionary{.}{%
}{.}\hspace{.4pt}1109\discretionary{/}{%
}{/}TVCG\hspace{.1pt}\discretionary{.}{%
}{.}\hspace{.4pt}2019\hspace{.1pt}\discretionary{.}{%
}{.}\hspace{.4pt}2934285}}


\bibitem{zhang_HowData_2020}
\href{https://doi.org/10.1145/3392826}{A.~X. Zhang, M.~Muller, and D.~Wang}.
\newblock \href{https://doi.org/10.1145/3392826}{How do data science workers
  collaborate? roles, workflows, and tools}.
\newblock \href{https://doi.org/10.1145/3392826}{{\em Proc. ACM Hum.-Comput.
  Interact.}}, \href{https://doi.org/10.1145/3392826}{4(CSCW1)},
  \href{https://doi.org/10.1145/3392826}{May 2020}.
  \href{https://doi.org/10.1145/3392826}
{doi: \textsf{%
10\hspace{.1pt}\discretionary{.}{%
}{.}\hspace{.4pt}1145\discretionary{/}{%
}{/}3392826}}


\end{thebibliography}
\end{document}